\documentclass[pre,twocolumn,showpacs,preprintnumbers,amsmath,amssymb,floatfix]{revtex4-1}
\usepackage{mciteplus}

\usepackage[dvips]{graphicx}    
\usepackage[normalem]{ulem}       
\usepackage{natbib}
\usepackage[usenames]{color} 

\newcommand{\be}{\begin{equation}}      
\newcommand{\bea}{\begin{eqnarray}}
\newcommand{\ee}{\end{equation}}
\newcommand{\eea}{\end{eqnarray}}
\newcommand{\bdm}{\begin{displaymath}}
\newcommand{\edm}{\end{displaymath}}
\newcommand{\half}{{\textstyle \frac{1}{2}}}    

\newcommand{\ten}[1]{\uuline{#1}{}}     

\newcommand{\C}{a_t}
\renewcommand{\vec}[1]{{\mathbf #1}}        

\begin{document}

\title{Negative Poisson's ratio and semi-soft elasticity of
  smectic-$C$ liquid crystal elastomers}

\author{A. W. Brown and J. M. Adams} 

\affiliation{SEPnet and the Department of Physics, Faculty of
  Engineering and Physical Sciences, University of Surrey, Guildford,
  GU2 7XH, United Kingdom.}

\begin{abstract}
  Models of smectic-$C$ liquid crystal elastomers predict that it can
  display soft elasticity, in which the shape of the elastomer changes
  at no energy cost. The amplitude of the soft mode and the
  accompanying shears are dependent on the orientation of the layer
  normal and the director with respect to the stretch axis. We
  demonstrate that in some geometries the director is forced to rotate
  perpendicular to the stretch axis, causing lateral expansion of the
  sample; a negative Poisson's ratio. Current models do not include
  the effect of imperfections that must be present in the physical
  sample. We investigate the effect of a simple model of these
  imperfections on the soft modes in monodomain smectic-$C$ elastomers
  in a variety of geometries. When stretching parallel to the layer
  normal (with imposed \emph{strain}) the elastomer has a negative
  stiffness once the director starts to rotate. We show that this is a
  result of the negative Poisson's ratio in this geometry through a
  simple scalar model.
\end{abstract}
\maketitle


\section{Introduction}

Liquid crystal elastomers (LCEs) are soft solids composed of flexible
polymers, with attached liquid crystalline mesogens, crosslinked into
a network \cite{warnerterentjev2007}. A variety of liquid crystalline
phases of LCEs have been synthesized, including the nematic and
smectic phases. The nematic phase undergoes deformation at no energy
cost, known as soft elasticity \cite{warnerbladon1994} in both
monodomain samples stretched perpendicular to the director
\cite{kundlerfinkelmann1995}, and in some types of polydomain samples
\cite{clarketerentjev1998, biggins:037802,doi:10.1021/ma0623688}. Soft
elasticity in nematic elastomers requires several sympathetic shears
to develop during the deformation as the director rotates. The
deformation of the sample must also obey the boundary conditions
imposed by the clamps. Thus, the sample forms a striped microstructure
on the micrometre length scale consisting of domains in which the
director rotates in opposite directions in adjacent domains
\cite{finkelmannkundler1997}. Without this microstructure soft
deformations would not be possible. In nematics this microstructure
has been observed in some detail experimentally \cite{zubarev1999},
and its mathematical properties described
\cite{desimonedolzmann2002}. Theoretically an ideal nematic LCE should
be perfectly soft, however in practise a small force must be applied
to deform the LCE. This semi-soft behaviour is due to various
imperfections in the elastomer, and can be incorporated into
theoretical models by the addition of a semi-soft energy term that
penalises rotation of the director with respect to the rubber matrix
\cite{PhysRevE.78.041704}.

\begin{figure}[!htb]
\begin{center}
\includegraphics[width = 0.48\textwidth]{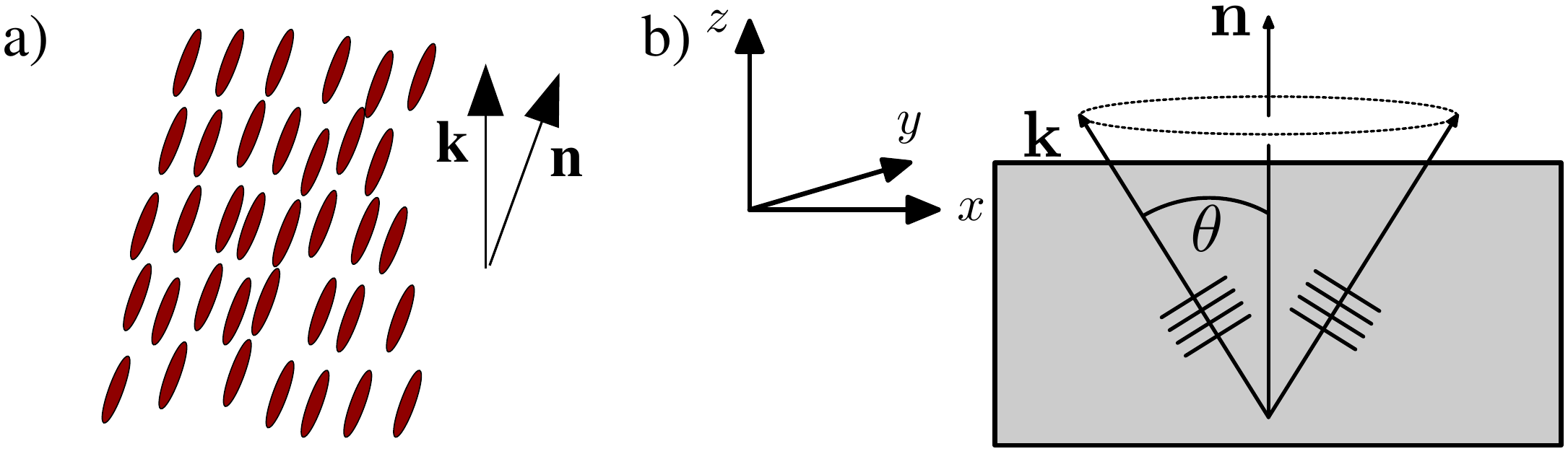}
\end{center}
\caption{a) The layer normal and director in the Sm-$C$ phase (polymer
  chains not shown), b) the director and layer normals in a
  pseudo-monodomain.}
\label{fig:pseudomd}
\end{figure}

Smectic LCEs have been fabricated in both the smectic-$A$ (Sm-$A$) and
smectic-$C$ (Sm-$C$) phases, and with both main chain
\cite{ferrer2008} and side chain \cite{nishikawa1999}
architectures. Their mechanical behaviour can be modelled by adding in
the embedded smectic layers to the nematic elasticity free energy
\cite{adams:05,stenull:011706}. Sm-$A$ elastomers with a high degree
of smectic order exhibit a sharp change in their elastic deformation
when deformed parallel to the layer normal, and are extremely
anisotropic materials, behaving as 2-D elastic materials
\cite{nishikawa1997,nishikawa1999,PhysRevE.82.031705}. However, the
response of smectic elastomers depends on the chemistry, the
crosslinking procedure, and domain sizes in the sample
\cite{PhysRevE.77.021706}. In contrast to the nematic phase the Sm-$A$
phase does not show any soft elastic behaviour because the director is
locked parallel to the layer normal. However, the elastic behaviour of
Sm-$C$ elastomers is predicted to be more complex. The director is
free to rotate on a cone around the layer normal, with fixed tilt
angle as shown fig.~\ref{fig:pseudomd} a). As a consequence it is
predicted to have a soft elastic mode just as in nematic elastomers
\cite{PhysRevE.72.011703,stenull:051709}. A more complicated
combination of shears is required in Sm-$C$ soft modes. As a result of
the compatibility requirements between these deformations a far more
restricted set of tensile geometries are predicted to deform softly
with clamped boundary conditions \cite{adams:07}.

To test these theoretical results experimentally, a monodomain must be
produced which requires alignment of both the layer normal and the
director. Using a two-stage crosslinking method, the director field
can be uniformly aligned \cite{MARC:MARC030121211}. However, the layer
normals are tilted at a fixed angle on a cone around the director.
We will refer to this as a \emph{pseudo-monodomain} (see
fig.~\ref{fig:pseudomd} b)) \cite{heinzefinkelmann2010} in order to
distinguish it from a monodomain, which has both the layer normal and
the director uniformly aligned, and polydomain, which has random
director and layer normal alignment. The layer normals in the
pseudo-monodomain can be aligned by a second uniaxial deformation
perpendicular to one of the layer normals \cite{semmler:95}, or
alternatively by a shear deformation perpendicular to the director
\cite{hiraoka:01}. To our knowledge no mechanical experiments on
monodomain Sm-$C$ elastomers have been reported, perhaps because of
the difficulty in aligning these samples. However, the spontaneous
deformations associated with changing the phase of the elastomer from
Sm-$A$ to Sm-$C$ have been observed \cite{hiraoka:05}. Many more
experiments have been carried out on the more accessible polydomain
system \cite{ferrer2008,sanchezferrer2011}. Unfortunately, as with
many polydomain systems, this is more difficult to model
theoretically.

The director reorientation in soft modes may be particularly important
for the electromechanical properties of chiral Sm-$C^*$ phase
elastomers \cite{adams:061704}. The liquid crystal rods have a
permanent dipole moment that is aligned perpendicular to both the
director, and the layer normal. The coupling between the macroscopic
mechanical deformations, and the microscopic orientation of these
dipoles results in their piezoelectric properties. These materials
show both the direct piezoelectric effect in pseudo-monodomains
\cite{2010papadopoulos}, as well as the inverse piezoelectric effect
\cite{hona2011}. Spontaneous polarization of pseudo-monodomains has
also been reported \cite{heinzefinkelmann2010}.

This paper is organised as follows. We will describe the model of
Sm-$C$ elastomers that will be use in \S \ref{sec:model}, and show
that the soft modes in this model have negative incremental Poisson's
ratio in some geometries. In \S \ref{sec:elongations} we will
illustrate the effect of the semi-soft elastic term in four different
geometries. We will then summarise the effect of this term, and
discuss the model predictions in relation to the mechanical
experiments in polydomains in \S \ref{sec:discussion}.

\section{Model Free energy}
\label{sec:model}

The model of a Sm-$C$ elastomer that will be used here is described in
Refs. \cite{adams:031706, stenull:021705}. The free energy has
contributions from the nematic elasticity $F_{nem}$, smectic layer
spacing $F_{sm}$, and the energy penalty for changing the tilt of the
director with respect to the layer normal $F_{tilt}$. The nematic
elasticity is given by
\begin{equation}
F_{nem} = \half \mu \mathrm{Tr} \left[ \ten{\lambda} \cdot \ten{\ell}_0
  \cdot \ten{\lambda}^T \cdot \ten{\ell}^{-1} \right]
\label{eqn:nem}
\end{equation}
where $\mu$ is the rubber shear modulus, and $\ten{\lambda}$ is the
deformation gradient. The step length tensor before the deformation
has been applied is $\ten{\ell}_0 = \ten{\delta} + (r-1) \vec{n}_0
\vec{n}_0$, with $\vec{n}_0$ the initial director, $\ten{\delta}$ the
unit tensor, and $r$ the polymer anisotropy. The current step length
tensor is denoted by $\ten{\ell}$, and its inverse by $\ten{\ell}^{-1}
= \ten{\delta} + (1/r-1) \vec{n} \vec{n}$, with $\vec{n}$ the final
director. In principle a Sm-$C$ elastomer should have a biaxial shape
tensor for the polymer backbone because its shape may be affected by
both the director alignment and the layer normal direction. For
simplicity we will approximate it as uniaxial here, depending only on
the director. We will also assume that the nematic and smectic order
parameters remain fixed throughout the deformation.

It is assumed that the smectic layers are embedded in the rubber
matrix, so that the corresponding layer normals $\vec{k}$ will deform
like embedded planes
\begin{equation}
  \vec{k} = \frac{\ten{\lambda}^{-T} \cdot \vec{k}_0}{|\ten{\lambda}^{-T} \cdot \vec{k}_0|}
\label{eqn:ln}
\end{equation}
where $\vec{k}_0$ is the initial layer normal. The layer spacing is
penalised by the smectic liquid crystal modulus $B$
\begin{equation}
F_{sm} = \half B \left(\frac{d}{d_0}-\frac{\cos \theta}{\cos \theta_0} \right)^2 
\label{eqn:sm}
\end{equation}
where $d$ is the final layer spacing, and $d_0$ is the initial layer
spacing, and $d/d_0 = 1/|\ten{\lambda}^{-T}\cdot
\vec{k}_0|$. $F_{sm}$ describes the free energy penalty for deviations
of the layer spacing away from that required to accommodate the
smectic mesogens. For tilted smectic mesogens, the required layer
spacing is $\cos \theta / \cos \theta_0$, where the their tilt angle
with respect to the layer normal is $\theta_0$ in the initial state,
and $\theta$ in the current state. The free energy term that penalises
the deviation of the director from a tilt angle $\theta_0$ is
\begin{equation}
F_{tilt} =\half \C \left(\cos^2 \theta_0 - (\vec{n}\cdot\vec{k})^2\right)^2,
\label{eqn:tilt}
\end{equation}
where $\C$ is the tilt modulus, and $\vec{n}\cdot \vec{k} = \cos
\theta$.

It will be assumed here that the bulk modulus of the rubber is much
larger that the shear, tilt and smectic moduli, so that the
deformation gradient obeys \mbox{$\mathrm{det}[ \ten{\lambda} ] = 1$},
hence it conserves volume. Typically the smectic layer modulus is very
large compared to the rubber shear modulus, i.e. $B\gg\mu$ (at least
in smectic elastomers of a similar type to that of Nishikawa et
al. \cite{nishikawa1997}), so that the layer spacing remains almost
fixed. The tilt modulus is also large compared to the shear modulus
$\C\gg \mu$, so that the tilt angle remains close to $\theta_0$
\cite{kramer:021704}.

\subsection{Soft elasticity}
\label{sec:softmodes}
The free energy outlined above permits the subset of the nematic soft
modes that maintain the layer spacing. There is only one soft mode
that satisfies this (up to a global rotation), and it corresponds to a
rotation of the director about the layer normal
\cite{PhysRevE.72.011703, stenull:051709}. We summarise some of the
properties of this soft mode here, as they are crucial in
understanding the semi-soft response of the elastomer.

\begin{figure}[!htb]
\begin{center}
\includegraphics[width = 0.48\textwidth]{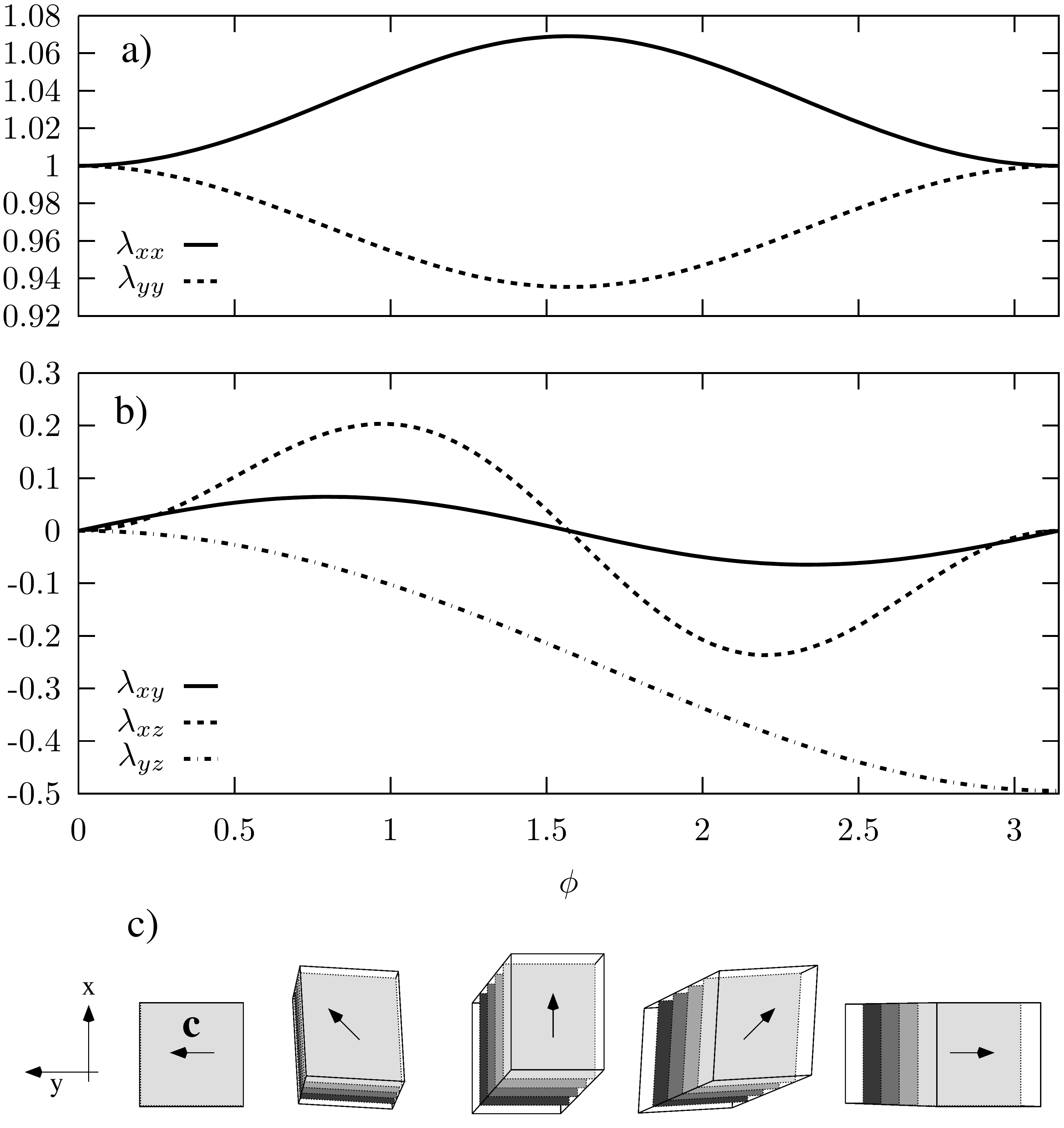}
\end{center}
\caption{For the parameter values $r = 2$ and $\theta_0 = 0.5$ radians
  a) shows the diagonal components of the deformation matrix, b) shows
  the shear components, and c) shows an illustration of the
  deformations on the LCE, together with the component of the director
  perpendicular to the layer normal, $\vec{c}$.}
\label{fig:smclambdas}
\end{figure}

We assume that the layer normal points along the $\vec{z}$ direction,
and the director is tilted into the $\vec{y}$ direction, i.e.
$\vec{k}_0 = \vec{z}$ and \mbox{$\vec{n}_0 = \vec{z} \cos \theta_0 +
  \vec{y} \sin\theta_0$} in the starting state. The soft modes can be
parameterised by the angle $\phi$ which gives the rotation of the
director $\vec{n}_0$ around the layer normal towards the $\vec{x}$
direction. The deformation matrix is given by
\cite{PhysRevE.72.011703}
\begin{equation}
\left(\!\!\begin{array}{ccc}
\frac{1}{a(\phi)} & \left( 1-\frac{\rho}{r}\right) \!\!\frac{\sin 2 \phi}{2 a(\phi)}&\frac{(r-1)\sin 2 \theta_0}{2 \rho}\!\left( \sin \phi-\left(1-\frac{\rho}{r}\right)\frac{\sin 2 \phi}{2 a(\phi)}\right)\\
0&a(\phi)  &\frac{(r-1)}{2 \rho}\sin 2 \theta_0 (-a(\phi) + \cos \phi)\\
0&0&1
\end{array}\!\! \right)
\label{eqn:softmode}
\end{equation}
where
\begin{eqnarray}
\rho &=&\sin ^2 \theta_0 + r \cos^2 \theta_0\\
a(\phi) &= &\sqrt{\cos ^2 \phi + \frac{\rho}{r} \sin^2 \phi}.
\end{eqnarray}
The deformation components as a function of rotation angle are
illustrated in Fig.~\ref{fig:smclambdas}.

\begin{figure}[!htb]
\begin{center}
  \includegraphics[width =
  0.48\textwidth]{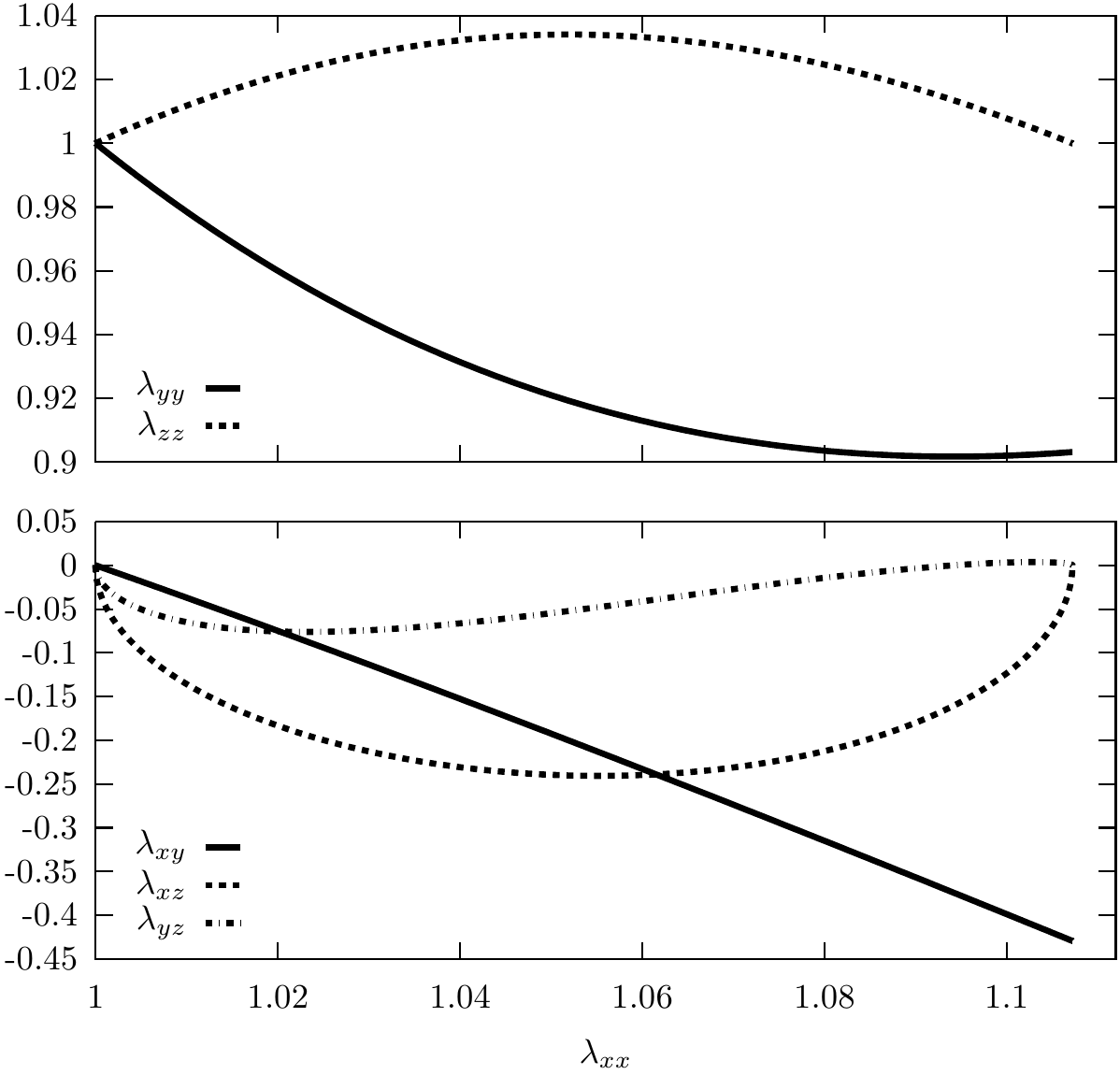}
\end{center}
\caption{The components of the upper triangular deformation matrix for
  the Sm-$C$ soft mode stretching parallel to the layer normal,
  with $r=2$ and $\theta_0 = 0.5$ radians. Initially $\vec{k}_0 =
  \vec{x}$ and $\vec{n}_0 = \cos \theta_0 \vec{x} + \sin \theta_0
  \vec{y}$.}
\label{fig:smlayernormal}
\end{figure}

\begin{figure*}[!htb]
\begin{center}
\includegraphics[width = 0.9\textwidth]{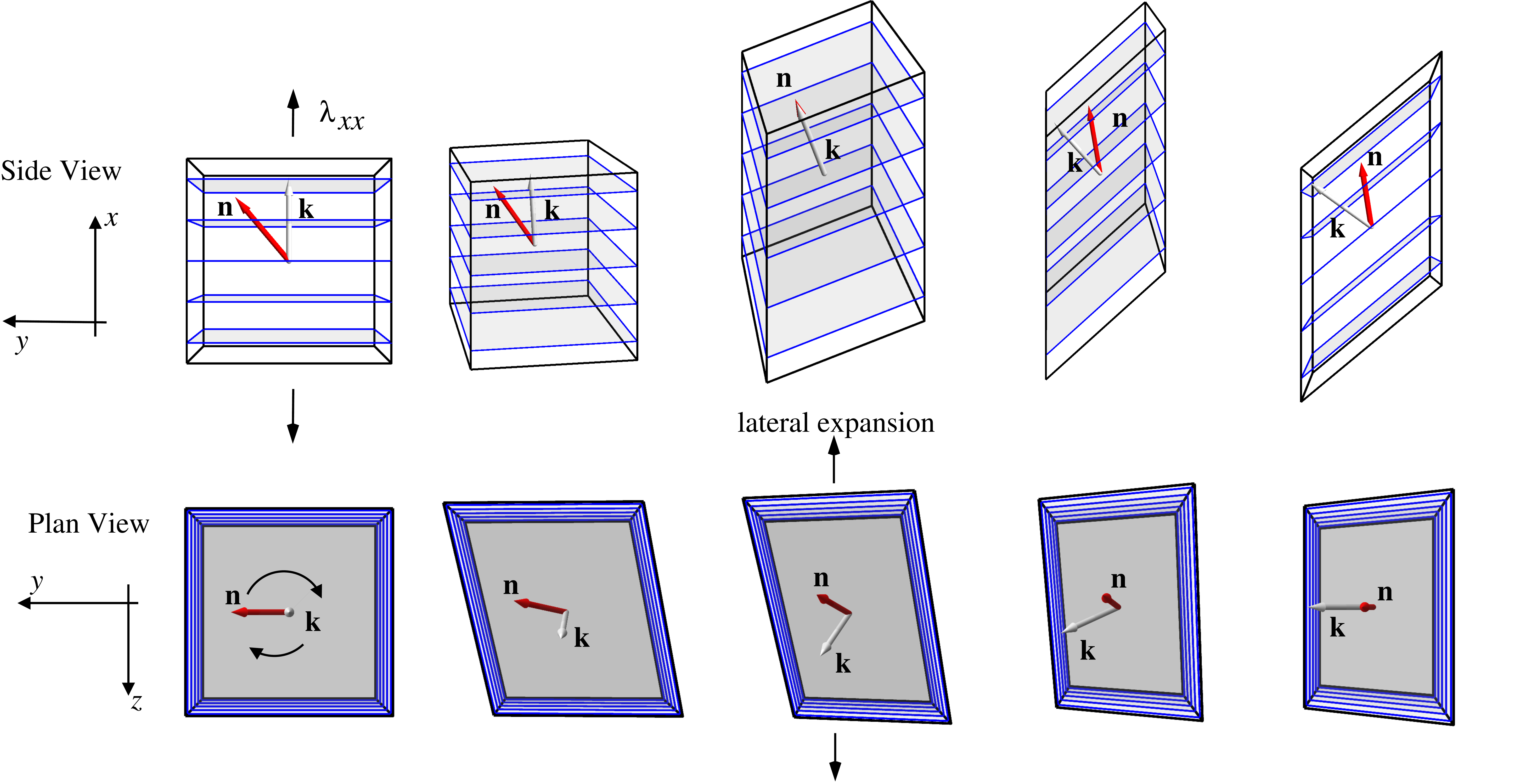}
\end{center}
\caption{An illustration of the Sm-$C$ elastomer deformation when
  stretching parallel to the layer normal. The director (red) moves
  out into the $z$ direction perpendicular to the stretch axis,
  maintaining its tilt angle with respect to the layer normal (white)
  causing the elastomer to expand in the perpendicular direction.}
\label{fig:illusln}
\end{figure*}

The soft mode in Eq.~(\ref{eqn:softmode}), denoted by
$\ten{\lambda}_\textrm{soft}$ can be transformed to different starting
configurations of the director and layer normal by the following
rotations
\begin{equation}
  \ten{\lambda}_\textrm{soft}^{\vec{k}_0} = \ten{P} \cdot\ten{Q}^T \cdot \ten{\lambda}^{\vec{z}}_\textrm{soft}(\phi) \cdot \ten{Q}
\end{equation}
where the rotation matrix $\ten{Q}$ takes the general starting layer
normal $\vec{k}_0$ to the $\vec{z}$ direction, and $\vec{n}_0$ into
$\cos \theta_0 \vec{z} + \vec{y} \sin \theta_0$. The second rotation
matrix $\ten{P}$ can be used to satisfy the requirements of the soft
mode in target state, for example ensuring that the $zx$ shear
component is zero. This transformation is described in Appendix
\ref{app:smlayernormal} for the case of stretching parallel to the
layer normal in the $\vec{k}_0 = \vec{x}$ direction, when $\vec{n}_0 =
\cos \theta_0 \vec{x} + \sin \theta_0 \vec{y}$. Although the result is
analytic, the algebra is not instructive, and is not presented
here. The components of the deformation matrix for this geometry are
illustrated in Fig.~\ref{fig:smlayernormal}.
The $\lambda_{zz}$ component increases with imposed $\lambda_{xx}$,
i.e. the sample expands in the direction perpendicular to the imposed
elongation. This is because the constraint requiring a fixed angle
between the layer normal and director results in the director rotating
into the $z$ direction. The sample then expands to accommodate the
anisotropic chain shape, as shown in Fig.~\ref{fig:illusln}.
This illustrates an unusual property of some Sm-$C$ soft modes: their
\emph{negative} Poisson's ratio. To our knowledge this mechanism for
negative Poisson's ratio has not been reported before. Alternative
mechanisms of producing auxetic behaviour based on modifying the
attachment of mesogens to the polymer backbone in smectic LCEs have
been proposed and investigated experimentally
\cite{wantingren,ren2009}. 

The incremental Poisson's ratio (IPR) is defined by
\begin{equation}
  \nu_{zz} = -\frac{d \lambda_{zz}}{d \lambda_{xx}}
\end{equation}
where an elongation $\lambda_{xx}$ is imposed and $\lambda_{zz}$ is
the transverse deformation. For isotropic materials, the Poisson's
ratio must be in the range $-1 < \nu < 0.5$. LCEs are anisotropic
materials, so have Poisson's ratios outside this range. As the
materials considered here are volume conserving, the Poisson's ratio
in the $y$ direction is $\nu_{yy} = 1-\nu_{zz}$. When stretching
parallel to the layer normal, the Poisson's ratio at $\lambda_{xx} =
1$ is given by
\begin{equation}
\label{eqn:IPRparak0}
  \nu_{zz} = - \frac{d \lambda_{zz}}{d\lambda_{xx}}\Big|_{\lambda_{xx} =
1} = -\frac{1}{(r - 1) \cos^2 \theta_0}
\end{equation}
Substituting in typical values of $\theta_0 \sim 0.5$ radians, and $r
\sim 2$ for a side chain system produces $\nu \sim -1.3$. This is
corresponds to a larger expansion than is achieved in auxetic foam
systems \cite{lakes1987}, albeit in only one direction. The extent of
the soft mode in this geometry is
\begin{equation}
\lambda_{xx}= \sqrt{1+\frac{(r-1)^2}{\rho^2} \sin^2 2
    \theta_0},
  \label{eqn:parak0amp}
\end{equation}
hence it has no extent when $\theta_0 = 0$, or when $r=1$.
Consequently the result in Eq.~(\ref{eqn:IPRparak0}) cannot be used to
calculate Poisson's ratio for the Sm-$A$ phase which has no soft
deformations.

\subsection{Semi-soft elasticity}

Soft modes in ideal LCEs have zero energy cost, and so the sample
requires no force to deform. In practise these materials have several
sources of non-ideal behaviour, such as compositional fluctuations and
cross linking points that result in semi-soft behaviour. We will use
the well known, and general form (up to quadratic order) of semi-soft
elasticity in nematics \cite{PhysRevE.78.041704}
\begin{equation}
  F_{ss} = \half \alpha \mu \rm{Tr}\left[\ten{\lambda}\cdot 
(\ten{\delta} - \vec{n}_0 \vec{n}_0^T)\cdot 
\ten{\lambda}^T \cdot \vec{n} \vec{n}^T \right].
\label{eqn:ss}
\end{equation}
Eq.~(\ref{eqn:ss}) is well founded in nematic LCEs, so serves as a
starting point for smectic LCEs. However, in Sm-$C$ elastomers the
semi-soft term in the free energy could in principle involve any of
the directions in the problem including the director, and the layer
normal, but we will neglect these effects here for consistency.

Typical values of $\alpha$ are up to $\sim 0.1$ in nematic LCEs, but
it may be even larger in smectic LCEs \cite{stenull:021705}.

Some studies of semi-soft elasticity have used the following
simplified form \cite{desimonedolzmann2002}
\begin{equation}
\label{eqn:nhss}
  F_{ss} = \half \alpha \mu \rm{Tr}\left[\ten{\lambda}\cdot 
    \ten{\lambda}^T \right],
\end{equation}
which is the neo-Hookean elasticity formula. This more general
semi-soft term gives rise to the same qualitative behaviour as
Eq.~(\ref{eqn:ss}).

\subsection{Numerical Method}
\label{sec:numerics}

The free energy described in Eq.~(\ref{eqn:nem}), (\ref{eqn:sm}),
(\ref{eqn:tilt}), and (\ref{eqn:ss}), is subject to the non-linear
constraints that the director remains of unit length and that the
layer normal deforms as an embedded plane (Eq.~\ref{eqn:ln}). This
constrained minimisation can only be performed analytically in a few
circumstances. Numerical minimisation of this free energy using
conventional methods often results in the location of only local
minima. We have used a simulated annealing algorithm to minimise the
total free energy, which finds the global free energy minimum more
reliably. The constraint of the tilt angle of $\theta$ between the
layer normal and the director can be encoded as
\begin{equation}
\vec{n} = \vec{c} \sin\theta + \vec{k} \cos \theta,
\end{equation}
where the vector $\vec{c}$ is perpendicular to $\vec{k}$. A particular
basis is required to express $\vec{c}$. It is convenient to use
$\vec{c}_0$, the starting component of $\vec{n}_0$ perpendicular to
$\vec{k}_0$, and $\vec{c}_0\times \vec{k}_0$. The vector $\vec{c}$ can
be expressed as
\begin{equation}
\vec{c} = \hat{\vec{a} }\cos\phi + \hat{\vec{b}} \sin \phi
\end{equation}
where $\hat{\vec{a}}$ is a unit vector constructed from the component
of $\vec{c}_0$ that is perpendicular to $\vec{k}$, and $\hat{\vec{b}}$
is perpendicular to both $\hat{\vec{a}}$ and $\vec{k}$,
i.e. $\hat{\vec{b}}=\vec{k} \times \hat{\vec{a}} $. The simulated
annealing algorithm then minimises the free energy over $\phi, \theta$
and the required components of $\ten{\lambda}$. The global minimum
derived from this was then refined using a \textsc{nag} sequential
quadratic programming library routine. The imposed constraints are
implemented using Lagrange multipliers. The results of this method are
in good agreement with the results obtained from configurations that
can be solved analytically.

\section{Elongations of Sm-$C$ elastomers}
\label{sec:elongations}
\begin{figure}[!htb]
\begin{center}
\includegraphics[width = 0.48\textwidth]{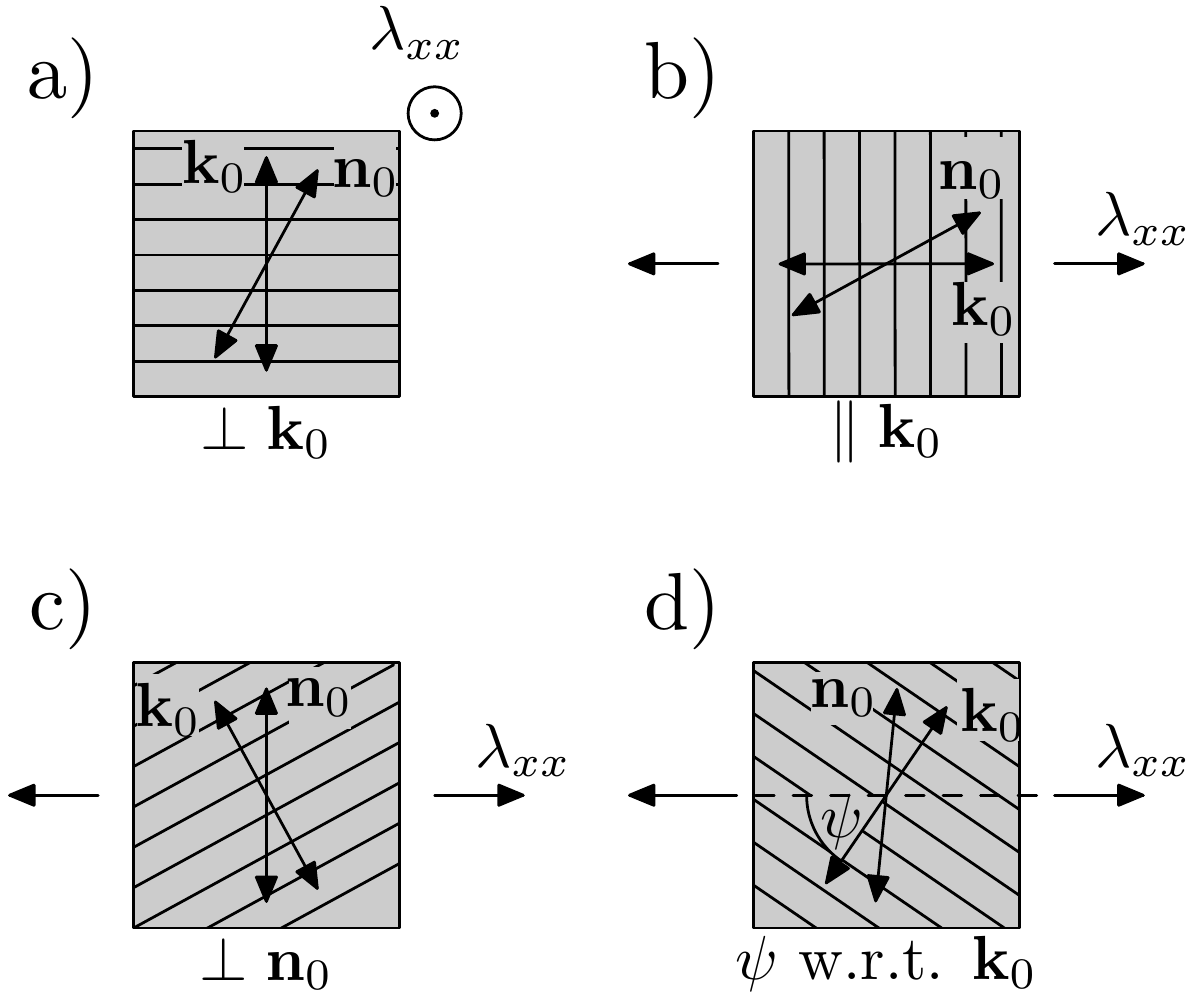}
\end{center}
\caption{The orientation of the director and layer normal for each of
  the elongations considered. a) perpendicular to the layer normal
  (note that the director and layer normal are both perpendicular to
  the $\vec{x}$ direction initially), b) parallel to the layer normal,
  c) perpendicular to the director and d) at an angle $\psi$ to the
  layer normal.}
\label{fig:geom}
\end{figure}

We will consider four elongations to illustrate some of the behaviour
and to build up some intuition for semi-soft Sm-$C$ elastomers. The
orientation of the layer normal and director in each case is shown in
fig.~\ref{fig:geom}.
The motivation for these different geometries is principally the
experimental work on polydomain Sm-$C$ elastomers \cite{ferrer2008},
and the mathematical studies of the Sm-$C$ free energy to find which
soft deformations are permitted by the formation of compatible
microstructures \cite{adams:07}. The elastic behaviour of the
polydomains is considerably more complicated. Understanding the
elastic deformations of a semi-soft monodomain is a useful step
towards modelling the experimentally more accessible polydomain
sample. We will ignore the effect of clamping at the boundaries, and
focus on the deformation of sheets of Sm-$C$ elastomer whose
mechanical properties will be dominated by the deformation of the
material in the middle of the long sheet. We will consider elongations
in the $\vec{x}$ direction, together with the induced shear
deformations. The appropriate deformation matrix is
\begin{equation}
\ten{\lambda} = \left( \begin{array}{ccc} 
\lambda_{xx} & \lambda_{xy} & \lambda_{xz}\\
0&\lambda_{yy} & \lambda_{yz}\\
0&0& \lambda_{zz}
\end{array} \right).
\label{eqn:defform}
\end{equation}
The $yx$ and $zx$ components are set to zero as they would be resisted
by counter torques. The $zy$ component can be set to zero by allowing
suitable rotations about the $\vec{x}$ axis, along which the elastomer
is stretched. We will consider an imposed deformation
$\lambda_{xx}$. In experiment, imposed stress ensembles are often
used, which yield the same results when the stress-strain curve is
monotonic. However, some of the stress-strain curves calculated here
are non-monotonic, hence there are several strain values for a single
stress value. In this case there is a difference between the fixed
stress and fixed strain ensembles, and for fixed stress a Maxwell
construction must be used to determine the strain. This is described
in \cite{PhysRevE.77.021702}, and briefly in \S \ref{sec:discussion}.

The model has the parameters $\mu, \C, B, r$ and $\theta_0$. Typical,
$\theta_0 \sim 30^\circ$ \cite{heinzefinkelmann2010}, $B/\mu = b \sim
60$ in well ordered samples
\cite{nishikawa1997,nishikawa1999,PhysRevE.77.021706}, $\C/\mu = c
\agt 1$ and $\alpha \sim 0.1$ in smectics
\cite{stenull:021705,kramer:021704}, and $r \sim 2$ in side chain
liquid crystalline polymers \cite{warnerterentjev2007}. We will use
these parameter values to illustrate the behaviour of the model in
what follows.

\subsection{Elongation perpendicular to $\vec{n}_0$ and $\vec{k}_0$}
\label{sec:perpn0k0}
\begin{figure}[!htb]
\begin{center}
\includegraphics[width = 0.48\textwidth]{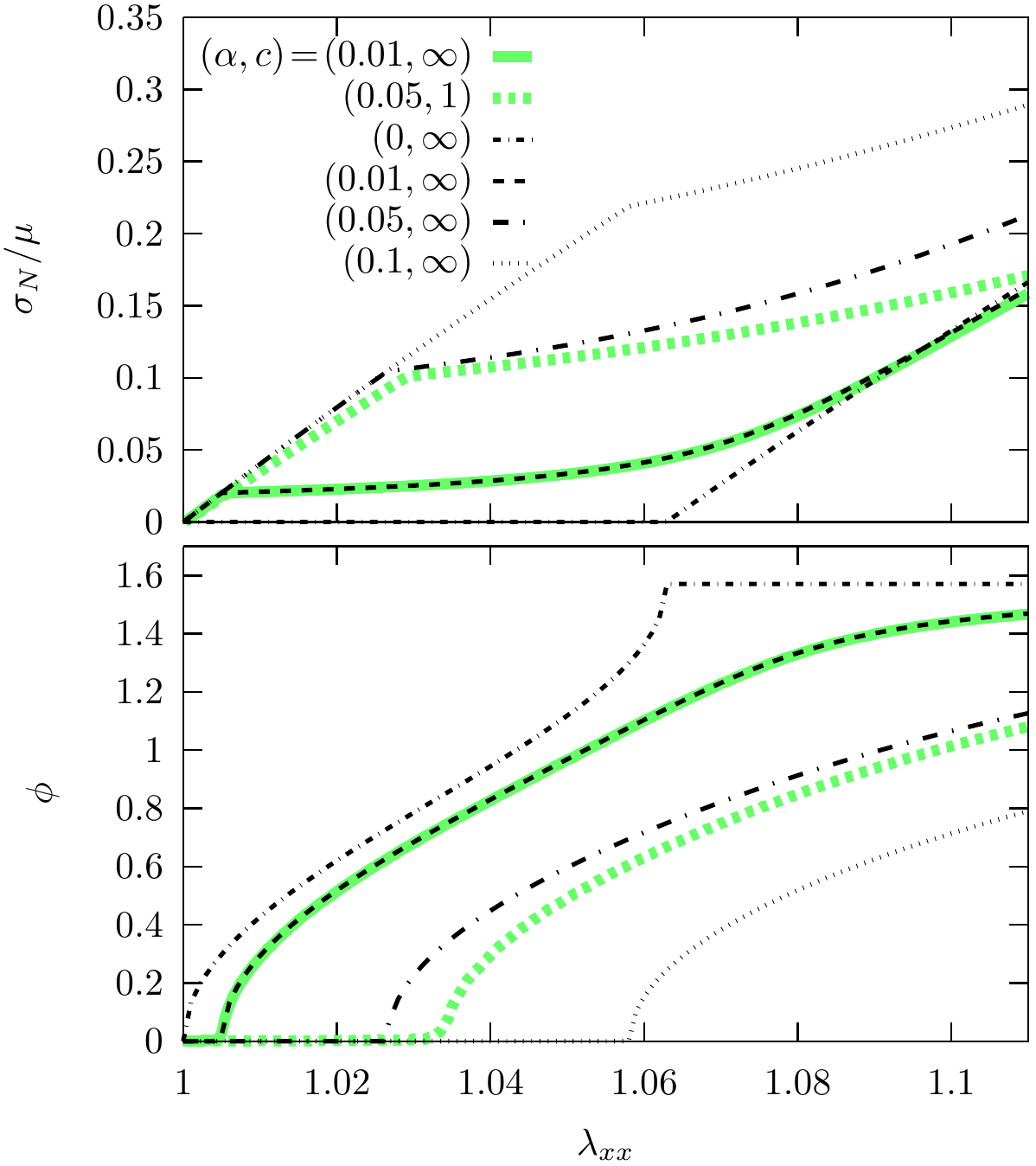}
\end{center}
\caption{The stress and the angle of rotation of a Sm-$C$ elastomer
  when stretched perpendicular to both the layer normal and the
  director. The model parameters are $b = 60$, $r=2$, $\theta_0 = 0.5$
  (radians) and the $\alpha$ and $c$ values shown on the figure. The
  thick curves (green) are from the more general numerical relaxation,
  and the black curves are calculated using the decomposition of the
  deformation matrix explained in the text.}
\label{fig:analsigperpk}
\end{figure}

First we consider an elongation deformation in the $\vec{x}$
direction, with the starting layer normal $\vec{k}_0=\vec{z}$ and the
starting director $\vec{n}_0 = \cos \theta_0\vec{z} + \sin\theta_0
\vec{y}$, as illustrated in fig.~\ref{fig:geom} a). In the absence of
the semi-soft term of Eq.~(\ref{eqn:ss}) this deformation is as
described in \S \ref{sec:softmodes}. The full free energy can be
minimised numerically as explained in \S \ref{sec:numerics}. The
resulting stress-strain curve, and the orientation of the director of
this minimisation are shown in fig.~\ref{fig:analsigperpk} by the
thick (green) lines.
For the ideal Sm-$C$ elastomer, this plateau ends at $\lambda_{xx} =
\sqrt{\frac{r}{\rho}}$, as can be seen from the soft mode in
Eq.~(\ref{eqn:softmode}). The plateau ends when the director has
completed a rotation by $\pi/2$ around the layer normal. For non-zero
values of $\alpha$ the onset of rotation of the layer normal is
delayed, and it never finishes a full $\pi/2$ rotation. This is
evident in the stress-strain curve, because the well defined stress
plateau for $\alpha=0$ becomes progressively less sharply defined.
For $\alpha\sim 0.01$ there is a pronounced stress plateau, but for
larger values of $\alpha\sim 0.1$ there is no plateau, merely a knee
in the stress-strain curve. Fig.~\ref{fig:analsigperpk} also shows the
effect of reducing the tilt modulus $c$. The knee in the stress strain
curve becomes less pronounced, and the rubber hardens more slowly for
larger values of $\lambda_{xx}$. The retardation of the director
rotation may be significant for piezoelectric response of these
materials. There would be no piezoelectric response until the strain
was above the threshold. The potential difference across the sample
would be lower in semi-soft samples because the alignment of the
electric dipoles associated with director rotation is spread over a
much larger deformation range.

\begin{figure}[!htb]
\begin{center}
\includegraphics[width = 0.48\textwidth]{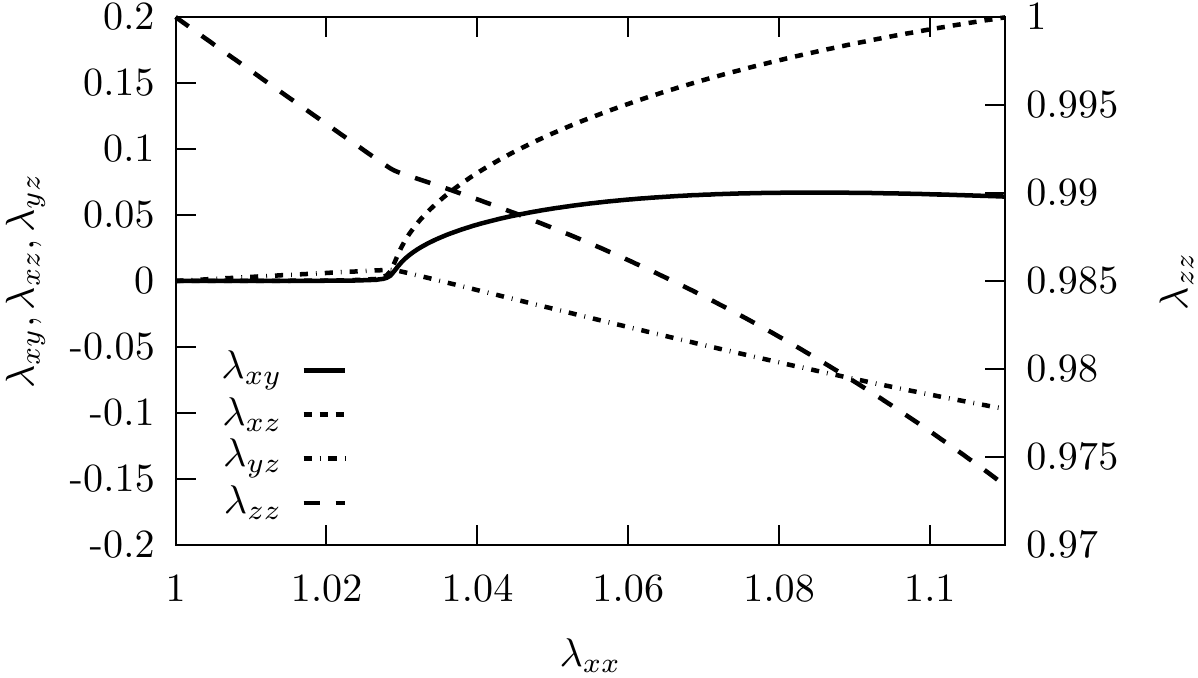}
\end{center}
\caption{The components of the deformation tensor for $(\alpha, b, c,
  \theta_0, r) = (0.05, 60, 1, 0.5, 2)$ when stretching perpendicular
  to both the director and the layer normal. Note that the sympathetic
  shears persist, as the director is unable to complete its $\pi/2$
  rotation.}
\label{fig:defsperpk}
\end{figure}

The deformation components when stretching perpendicular to $\vec{k}$
are illustrated in fig.~\ref{fig:defsperpk}.
Note the sympathetic shears that accompany the director rotation are
persistent, because the director rotation is never completed if
$\alpha >0$.

Numerically it is clear that with the inclusion of the semi-soft term
there is a delay in the rotation of the director. Some analytical
progress can be made in this geometry by decomposing the deformation
into three parts; the initial hard deformation with fixed director and
layer spacing denoted $\ten{\lambda}_\mathrm{hard}$, the soft mode
$\ten{\lambda}_\mathrm{soft}$ and the subsequent shear and elongation
after the soft mode $\ten{\lambda}^\prime$ \cite{PhysRevE.77.021702}
\begin{equation}
  \ten{\lambda} = \ten{\lambda}^\prime\cdot\ten{\lambda}_\mathrm{soft}\cdot\ten{\lambda}_\mathrm{hard}
\end{equation}
where $\ten{\lambda}_\mathrm{hard} = \mathrm{diag} (\lambda_1,
1/\lambda_1, 1)$, $\ten{\lambda}_\mathrm{soft}$ given in
Eq.~(\ref{eqn:softmode}), and
\begin{equation}
\ten{\lambda}^\prime = \left(
\begin{array}{ccc}
\zeta &0&\eta\\
0&1/\zeta & 0 \\
0&0&1
\end{array}
\right).
\end{equation}
This deformation matrix can be substituted into the free energy terms
of Eq.~(\ref{eqn:nem}), (\ref{eqn:sm}), and (\ref{eqn:ss}) (assuming
that $c\rightarrow \infty$, so that $\theta = \theta_0$). The problem
is then reduced to a minimisation over the variables $\lambda_1,
\zeta, \eta$ and $\phi$, with the constraint that the total
$\lambda_{xx}$ is prescribed. The threshold before the onset of
director rotation can be calculated by setting $\zeta = 1$ and $\eta =
0$, then performing a series expansion of the free energy in soft mode
rotation angle $\phi$. The leading term is $\mathcal{O}(\phi^2)$, and
when this term becomes negative a non-zero value of $\phi$ will lower
the free energy. To leading order in $(\lambda_1-1)$, this coefficient
becomes negative when $\lambda_1$ is approximately
\begin{eqnarray}
  \nonumber
  \lambda_1  && = 1+8 r^2 \alpha/(1 + 29 r - 29 r^2 - r^3 + r \alpha + 35 r^2 \alpha  \\
  &&+ 4 r^2 \alpha \cos
  2 \theta + (r-1  ) ( (r-1)^2+ r  \alpha) \cos
  4 \theta)
\end{eqnarray}
This value is slightly smaller than the corresponding threshold to
director rotation in nematic elastomers of $\lambda_1^3 =
\frac{r-1}{r-1-\alpha r}$ \cite{warnerterentjev2007}. Intuitively this
is because in the Sm-$C$ phase the deformation is restricted to two
dimensions by the layer spacing constraint. Consequently there is a
larger contraction in the direction perpendicular to the stretch which
causes the elastic free energy to rise faster, and hence the director
rotation to start earlier in Sm-$C$ LCEs as compared to the nematic
phase.

The minimisation of the free energy over $\lambda_1, \zeta, \eta$ and
$\phi$ produces results that are in good agreement with the more
general numerical method. The results are shown by the black lines in
Fig.~\ref{fig:analsigperpk}.

\subsection{Elongation parallel to $\vec{k}_0$}
\label{sec:parak0}

\begin{figure}[!htb]
\begin{center}
\includegraphics[width = 0.48\textwidth]{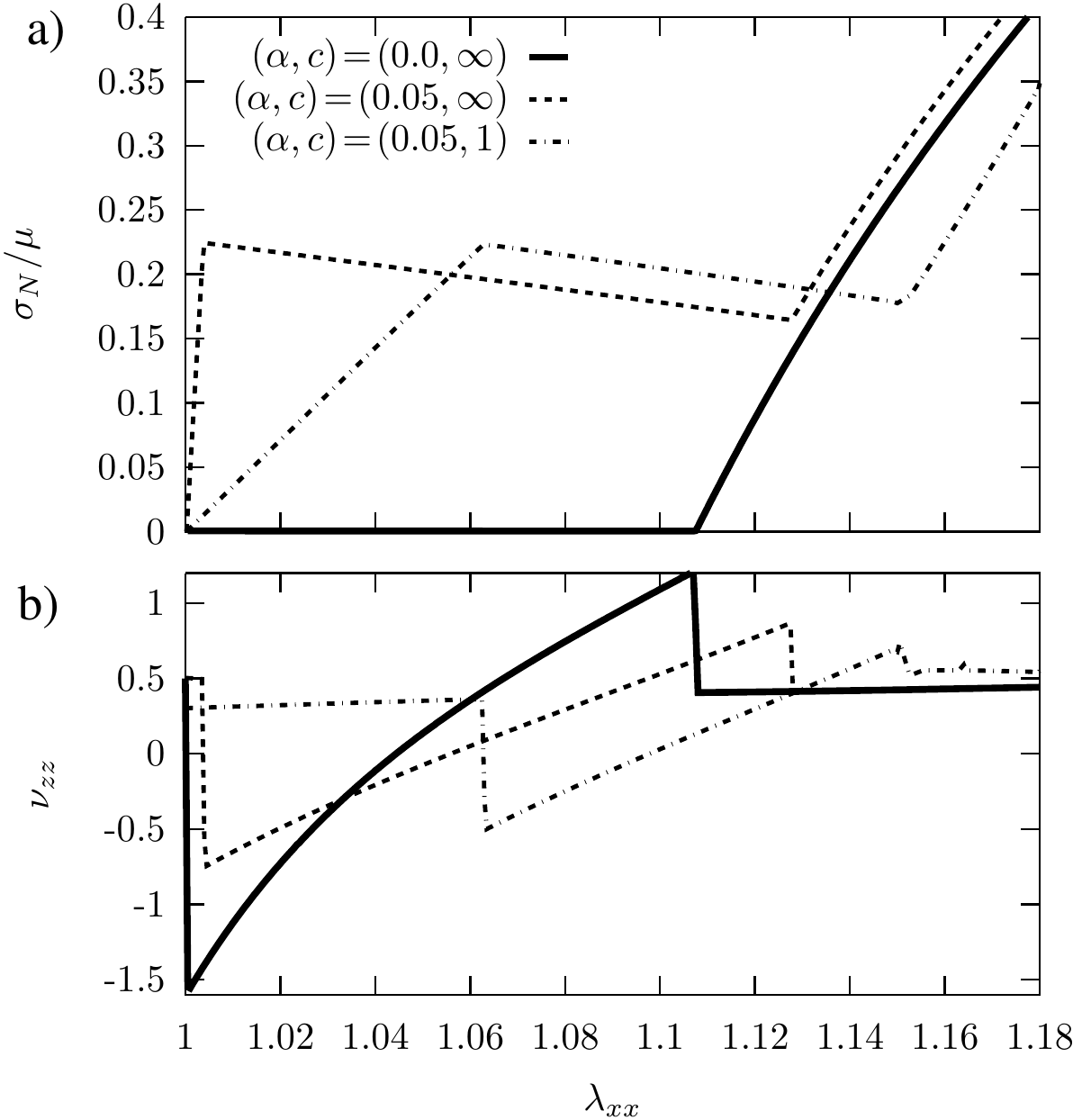}
\end{center}
\caption{a) The stress-strain response for a semi-soft Sm-$C$
  elastomer stretched parallel to the layer normal. The model
  parameters are $b=60$, $r=2$ and $\theta_0 = 0.5$, and the values of
  $(\alpha, c)$ shown in the figure. b) The corresponding IPRs for the
  stress-strain curves. }
\label{fig:stressparak0}
\end{figure}

Elongation parallel to the layer normal is illustrated in
Fig.~\ref{fig:geom} b). The initial layer normal is given by
$\vec{k}_0 = \vec{x}$ and the director $\vec{n}_0 = \vec{x} \cos
\theta_0 + \vec{z} \sin \theta_0$. Using the form of deformation
matrix described in Eq.~(\ref{eqn:defform}), the free energy can again
be minimised using the numerical technique described in \S
\ref{sec:numerics}. The results for various values of the semi-soft
parameter $\alpha$ are illustrated in fig.~\ref{fig:stressparak0}.
For $c\rightarrow \infty$ the first part of the stress-strain curve is
determined by the smectic layer modulus $B$. The semi-soft term
prevents the rotation of the director, and the layer spacing
increases. Once the force required to increase the layer spacing is
comparable to that required to rotate the director the semi-soft mode
begins. The stress-strain curve has negative slope once director
rotation starts. As explained in \S \ref{sec:softmodes} there is a
negative IPR in this geometry as the director rotates around the layer
normal into the direction perpendicular to the stretch axis (see
Fig.~\ref{fig:stressparak0} b)). This lateral expansion, combined free
energy expression for the semi-soft elasticity, results in the
negative stiffness. For larger values of $\alpha$ the Poisson's ratio
becomes less negative.

\begin{figure}[!htb]
\begin{center}
\includegraphics[width = 0.48\textwidth]{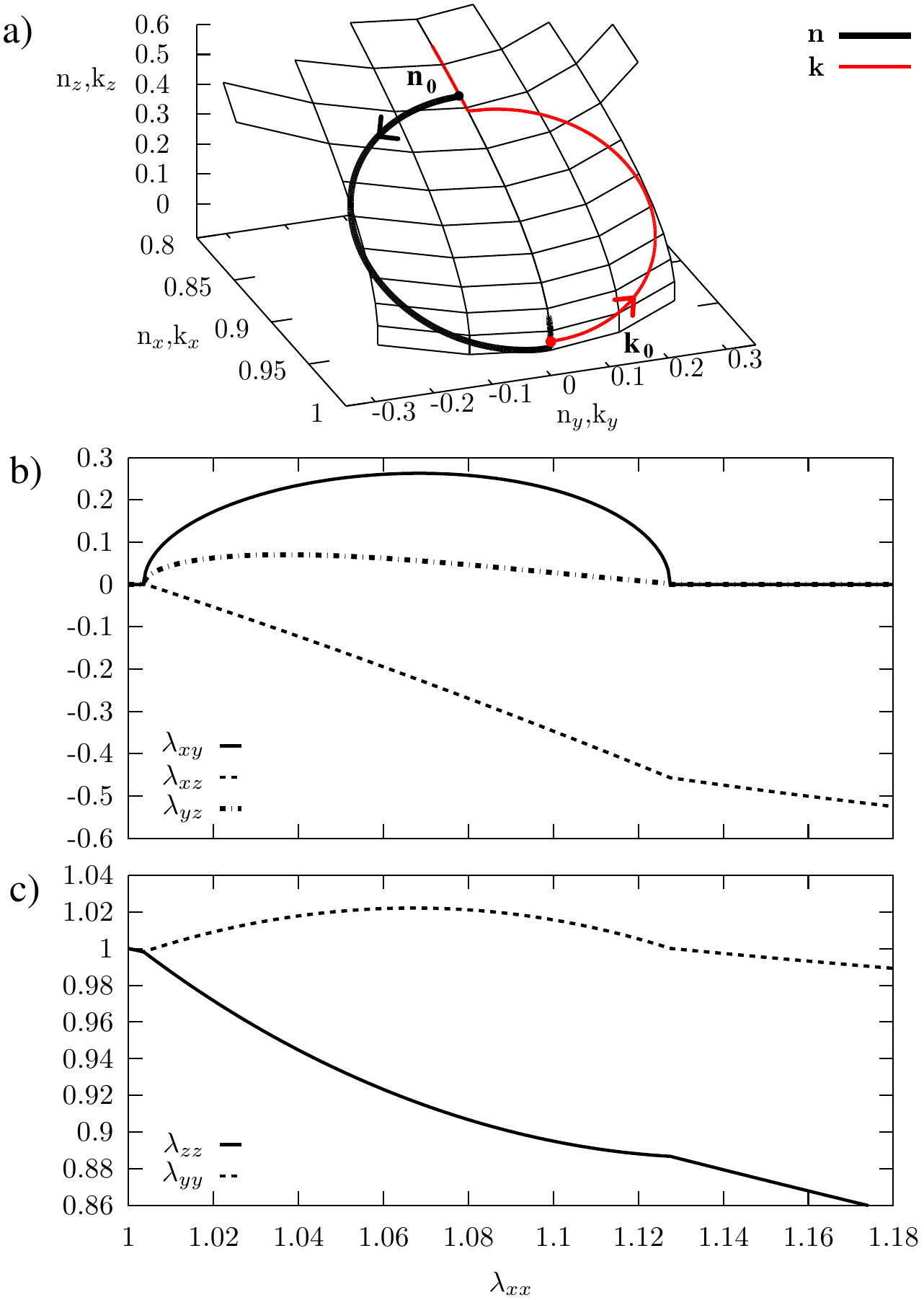}
\end{center}
\caption{For the parameter values $(\alpha, c) = (0.05, \infty)$ and
  $b=60$, $r=2$ and $\theta_0 = 0.5$ radians for stretching parallel
  to the layer normal. a) Shows the director and layer normal
  rotation,b) the shear components, and c) the diagonal components of
  the deformation tensor when stretching parallel to the layer normal.
}
\label{fig:parak0nkdef}
\end{figure}

The rotation of the layer normal and director, and the deformation
components are illustrated in Fig.~\ref{fig:parak0nkdef}. The
expansion of the sample in the $z$ direction is clearly visible at the
onset of rotation, as are the usual shear components that accompany a
soft mode. For finite values of $c$ the deformation becomes more
complicated; before the threshold the director rotates towards the
layer normal and the sample shears, which itself results in movement
of the layer normal. There is both an increase in the threshold to the
start of rotation, and a reduction in the amplitude of the semi-soft
deformation. This is because the shearing before director rotation
results in rotation of the layer normal, and there is a reduction in
the tilt angle before the onset of shearing.

The soft mode in this geometry can be calculated analytically, as
explained in \S \ref{sec:softmodes}, and its amplitude is given in
Eq.~(\ref{eqn:parak0amp}). Some analytic results can be obtained by
decomposing the deformation as follows.
\begin{equation}
  \ten{\lambda} = \ten{\lambda}^\prime\cdot\ten{\lambda}_\mathrm{soft}\cdot\ten{\lambda}_\mathrm{hard}.
\end{equation}
Both $\ten{\lambda}'$ and $\ten{\lambda}_\mathrm{hard}$ have the form
\begin{equation}
\left( \begin{array}{ccc}
\eta & 0& \zeta\\0& \xi & 0\\0&0&1/(\eta \xi)
\end{array}
\right).
\end{equation}
The appropriate soft mode must be calculated based on the rotated
layer normal, as the shear component $\zeta$ will cause it to rotate.

The onset of director rotation can be obtained by substituting back
into the free energy, expanding in terms of $\phi$ up to quadratic
order. When the coefficient of the $\mathcal{O}(\phi^2)$ term is
negative, the soft mode becomes active. This happens when
\begin{eqnarray}
\nonumber
\lambda_1 &&\approx 1+\frac{r^2 \alpha  }{b (r-1)^2 \cos^2 \theta_0}+ \frac{3 \alpha r^2 }{4 \rho^2 b^2 c \cos^2 \theta_0}\\
&&+\mathcal{O}\left(b^{-3},c^{-2}, \alpha^2 \right).
\end{eqnarray}
Note for smaller values of $c$ this is inaccurate because the shear is
only expanded up to quadratic order. The occurrence of $\alpha$ and $b$
in this expression correspond to the competition between the
stretching of the layer spacing, and the semi-soft elastic term
keeping the director fixed in the matrix. The threshold predicted by
this calculation is consistent with the numerical results for large
$b$ and $c$.

\subsubsection{Scalar model of negative slope region}

\begin{figure}[!htb]
\begin{center}
\includegraphics[width = 0.48\textwidth]{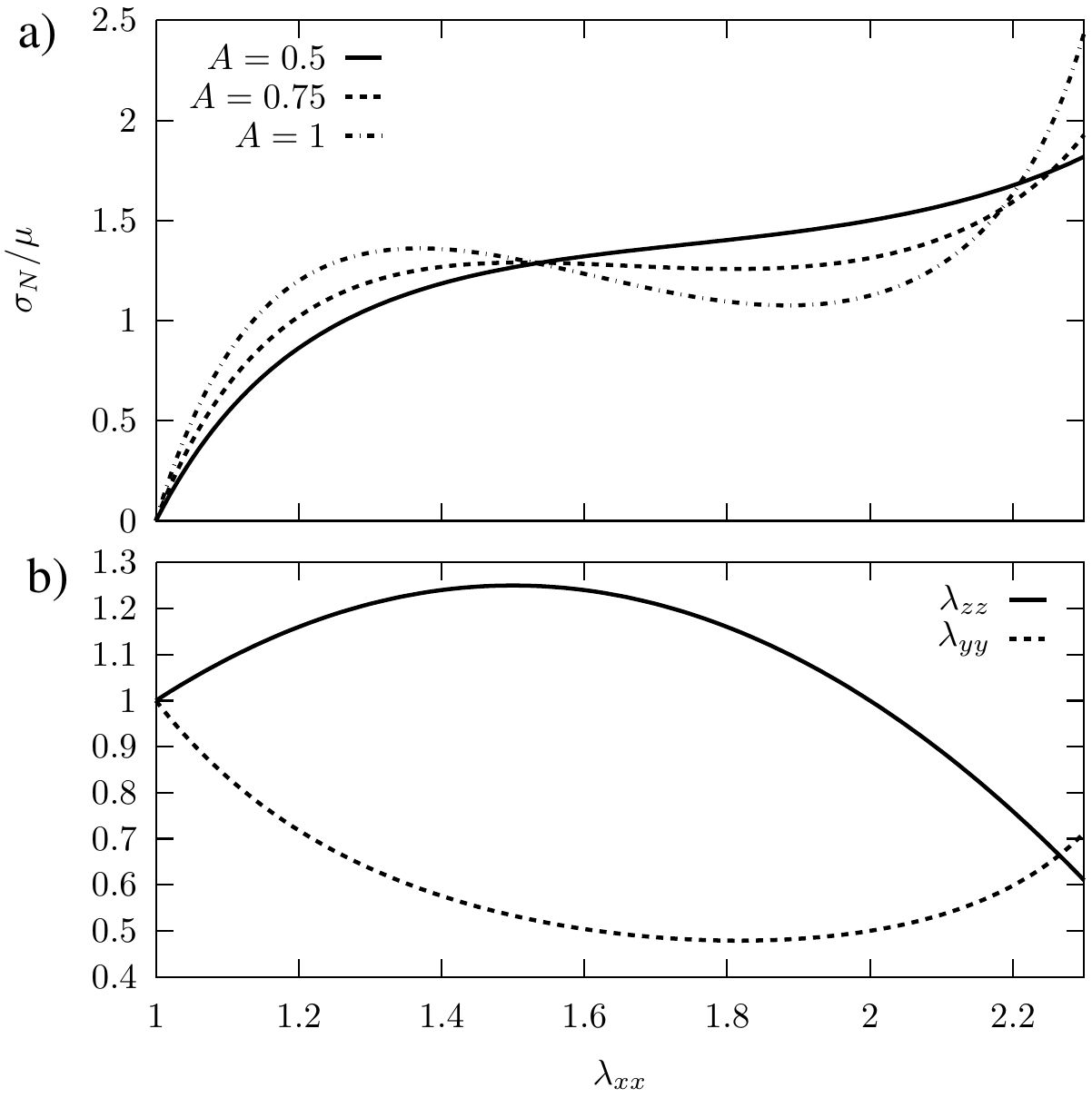}
\end{center}
\caption{For the scalar model of the negative stress strain curve
  described in the text a) shows the stress-strain curves for $A =
  0.5, 0.75, 1$, and b) the deformation components for $A = 1$ for the
  scalar auxetic model. }
\label{fig:toyaux}
\end{figure}

The unusual response above for the Sm-$C$ soft mode can be illustrated
for a much simpler deformation. Consider an elongation with a diagonal
deformation matrix of an imposed $\lambda_{xx}$, $\lambda_{zz}$ given
by
\begin{equation}
  \lambda_{zz} = 1-A \left(\lambda_{xx} - \frac{3}{2}\right)^2+\frac{A}{4},
\label{eqn:toy}
\end{equation}
with $\lambda_{yy}$ determined by volume conservation. The parameter
$A$ here controls the initial rate of expansion of the material. Its
Poisson's ratios are $-A$, and $1+A$. This is similar to the soft mode
in a Sm-$C$ illustrated in Fig.~\ref{fig:smlayernormal}. The
deformation in Eq.~(\ref{eqn:toy}) can be substituted into a
neo-hookean model such as Eq.~(\ref{eqn:nhss}), which is broadly
similar to the semi-soft elastic energy term. The resulting
stress-strain curve is shown in fig.~\ref{fig:toyaux}.
It can be seen from this plot that for sufficiently large values of
$A$ the stress-strain curve has a negative slope similar to stretching
the Sm-$C$ LCE parallel to the layer normal. For some geometries the
Poisson's ratio is sufficiently negative to result in a negative
stiffness. The configurational entropy of the perpendicular degrees of
freedom decreases as the sample expands resulting in a positive
contribution to the stress. Once lateral expansion starts to slow
sufficiently there is a weaker contribution to stiffness of the sample
from the perpendicular degrees of freedom and the stress starts to
drop, which produces a negative slope in the stress-strain
response. By tuning the parameter $A$ in the model, the balance
between the parallel and perpendicular degrees of freedom can be
altered, and the stiffness changed from negative to positive.

This scalar model shows that the negative stiffness is a result of the
lateral expansion during the Sm-$C$ soft mode, and not due to the form
of the semi-soft elastic term.

\subsection{Elongation perpendicular to $\vec{n}_0$}

\begin{figure}[!htb]
\begin{center}
\includegraphics[width = 0.48\textwidth]{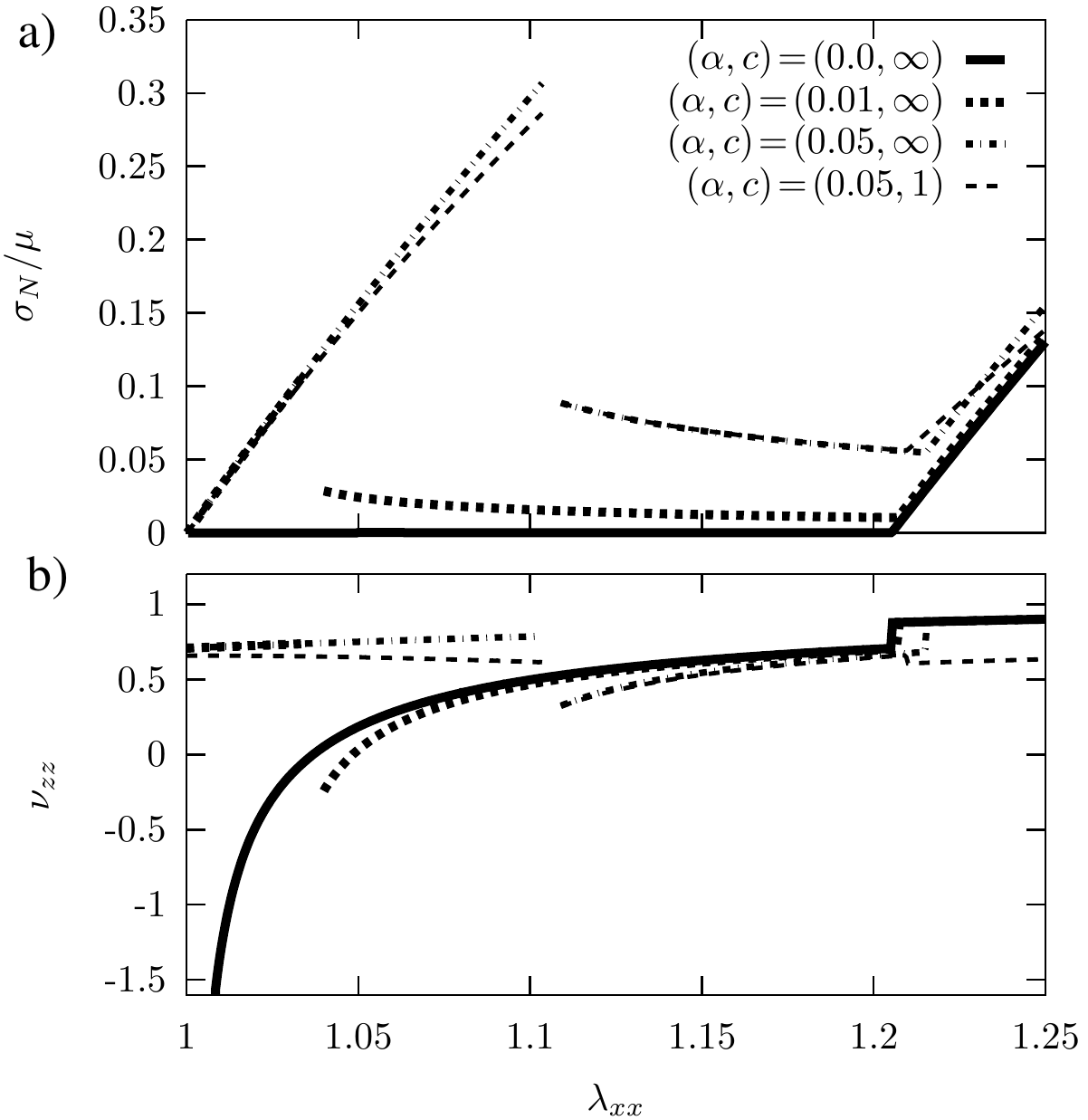}
\end{center}
\caption{a) The stress-strain curves for stretching perpendicular to
  the layer normal, b) the IPR for various parameters $b=60$, $r=2$
  and $\theta_0 = 0.5$ and values of $(\alpha, c)$ shown on the
  figure.}
\label{fig:ssperpn0}
\end{figure}

\begin{figure}[!htb]
\begin{center}
\includegraphics[width = 0.48\textwidth]{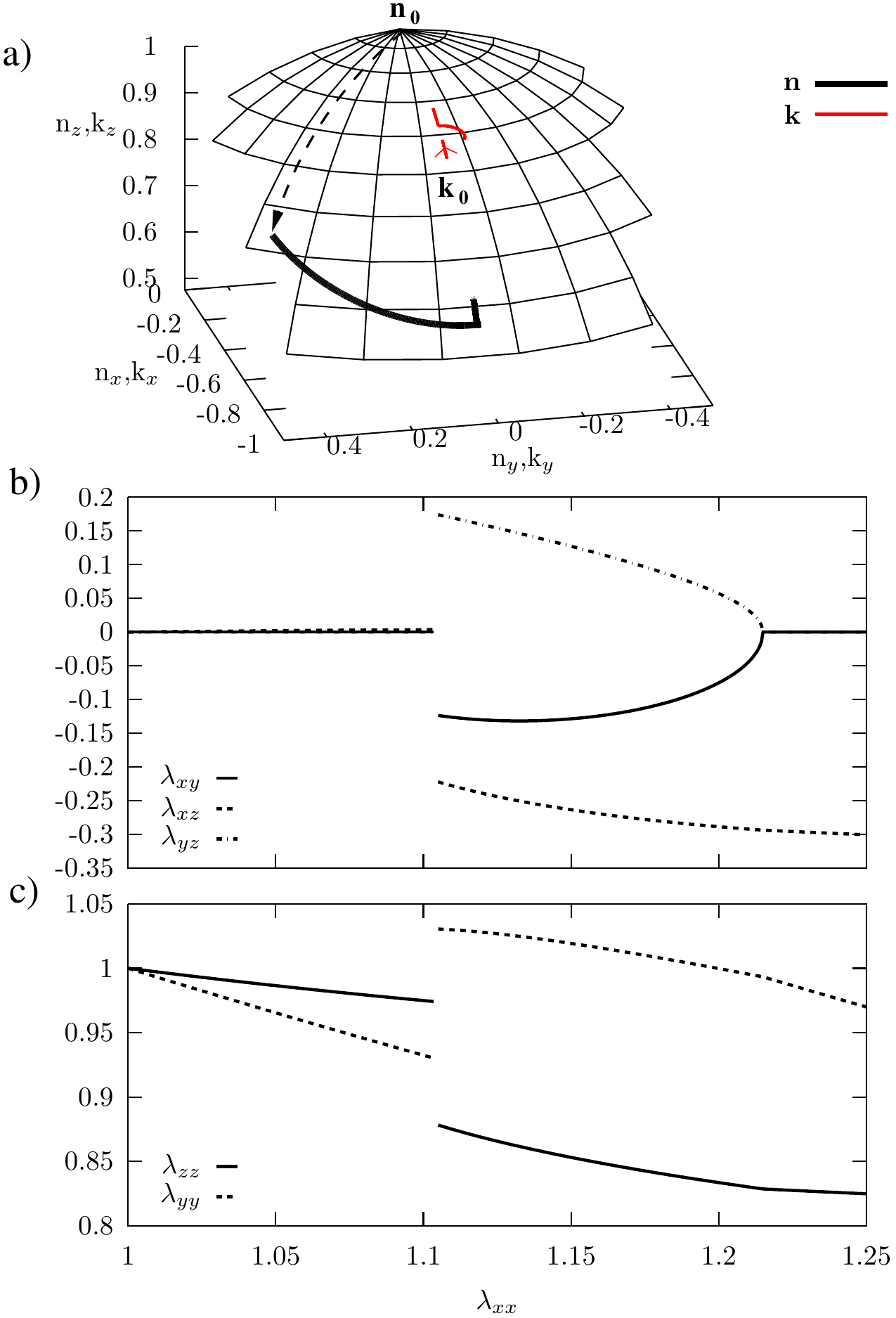}
\end{center}
\caption{When stretching perpendicular to the director, for the
  parameter values $(\alpha, c) = (0.05, \infty)$ and $b=60$, $r=2$
  and $\theta_0 = 0.5$ radians a) shows the director and layer normal
  rotation, b) the shear components, and c) the diagonal components of
  the deformation tensor when stretching perpendicular to
  $\vec{n}_0$. }
\label{fig:defperpn0}
\end{figure}

Stretching perpendicular to the initial director, $\vec{n}_0$ is
illustrated in fig.~\ref{fig:geom} c). The results for the numerical
calculation of the stress-strain curve for this geometry are shown in
Fig.~\ref{fig:ssperpn0}.
This geometry has the remarkable feature that $\nu_{zz} \rightarrow
-\infty$ when $\alpha \rightarrow 0$, as shown in
Fig.~\ref{fig:ssperpn0}. For larger values of $\alpha$ the Poisson's
ratio becomes less negative. The jump in the director also causes a
discontinuity in the IPR, and a sudden increase in the width of the
sample. Note that in this geometry there is a discontinuity in the
stress-strain curve, in addition to the negative stiffness. The
discontinuity in the stress-strain curve is accompanied by a jump in
the director as shown in Fig.~\ref{fig:defperpn0}.
Intuitively the discontinuity arises because when the director jumps
the long axis of the polymer shape tensor jumps towards the elongation
direction. Consequently the natural length of the rubber in this
direction is increased, so there is corresponding drop in the
stress. 

\begin{figure}[!htb]
\begin{center}
\includegraphics[width = 0.48\textwidth]{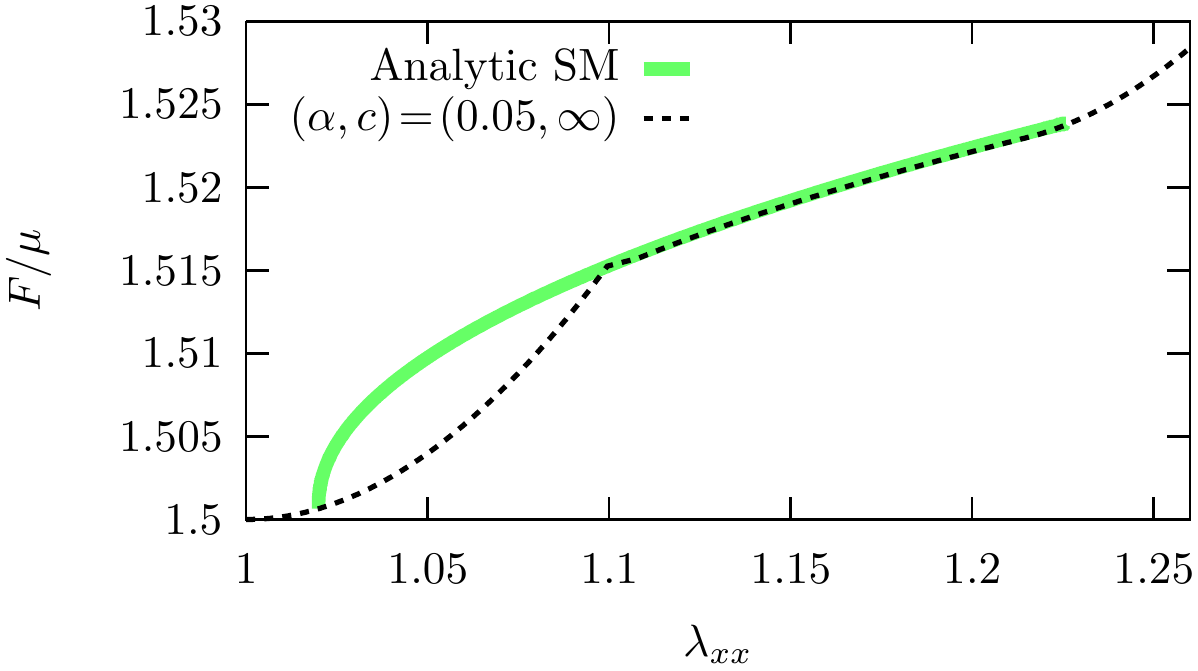}
\end{center}
\caption{The free energy calculated numerically, and the free energy
  trajectory of the semi-soft mode with continuous director rotation
  when stretching perpendicular to the director. Here $b=60, r=2,
  \theta_0 = 0.5$ and $(\alpha, c)$ are shown on the figure.}
\label{fig:perpn0sm}
\end{figure}

The jump in the director can be understood from the properties of the
soft mode in this geometry. We can approximate the first part of the
total deformation (until the end of director rotation) as a hard
deformation where there is no director rotation, followed by a soft
mode
\begin{equation}
\ten{\lambda} = \ten{\lambda}_\textrm{soft}\cdot\ten{\lambda}_\textrm{hard}.
\end{equation}
The soft mode in this geometry can be calculated analytically as
explained in appendix \ref{app:smlayernormal}. Whilst its analytic
form is algebraically very long, the amplitude of the soft mode has a
much simpler expression, and is given by
\begin{eqnarray}
\nonumber
  \lambda_{xx}&& = \left(3 + r(7 r - 2)+ 4(r^2-1)\cos 2 \theta_0+\right.\\
&&\left.(1+(2-3r)r)\cos 4\theta_0\right)^{1/2}/(2 \sqrt{2} \rho).
\end{eqnarray}
The hard part of the deformation has only diagonal elements, and an
$xz$ shear component.
\begin{equation}
\ten{\lambda}_\textrm{hard} = \left( 
\begin{array}{ccc}
\lambda_{xx} & 0 & \lambda_{xz} \\
0&1/(\lambda_{xx} \lambda_{zz}) & 0\\
0&0&\lambda_{zz}
\end{array}
\right)
\end{equation}
Substituting this into the full free energy density yields an
approximate solution to the minimisation problem, where the director
rotation is assumed to be continuous. The free energy density in this
case is shown in Fig.~\ref{fig:perpn0sm}.
The analytic solution with continuous director rotation has higher
free energy for the first part of the deformation. Hence, the
elastomer initially stretches without director rotation. If the
director were to start rotating, then the form of the soft mode
results in rapid rotation of the director, and an infinite slope in
the free energy. However, the rate of increase slows, and eventually
the state with a rotated director is lower in free energy than that
with a fixed director. At this point the director jumps to the new
orientation. There is a discontinuity in the slope of the free energy
at this point, or equivalently a jump in the stress.

This behaviour is not solely a result of the semi-soft energy term,
but again is a result of the shape of the soft mode, combined with a
general semi-soft elasticity term. These calculations are based on an
equilibrium model of a Sm-$C$ elastomer. In practise kinetic terms,
such as viscosity would smooth out the sharp jump demonstrated here.

\subsubsection{Scalar  model describing stress discontinuity}

The semi-soft behaviour of Sm-$C$ elastomers is characterised by two
deformation modes; before the onset of director rotation, and
afterwards. A scalar model that exhibits the same behaviour when
stretching perpendicular to the director can be developed based on
representing each of these deformation modes as a spring, and
deforming the two springs in series. The total strain is the sum of
two deformation modes corresponding to keeping a fixed director
$\epsilon_\textrm{U}$, and rotating the director $\epsilon_\textrm{SM}$
\begin{equation}
  \epsilon_\textrm{T} = \epsilon_\textrm{U}+\epsilon_\textrm{SM}.
\end{equation}
The two modes of deformation have different energy penalties, the
first arises from a simple uniaxial deformation, so in a neo-hookean
energy model will result in a free energy term of the form
\begin{equation}
F_\textrm{U} = \half K_1 \epsilon_\textrm{U}^2,
\end{equation}
where $K_1$ corresponds to the shear modulus of the rubber. The second
arises from the soft mode, which has a singular edge in the
contraction of the rubber as it is stretched. The $zz$ component in
the soft mode is initially of the form $\lambda_{zz} =
1/(1+(\lambda_{xx}-1)^\beta)$ (where here $\lambda_{xx} -1 =
\epsilon_\textrm{SM}$. When this is put into the neo-hookean free
energy, it results in free energy terms to leading order in
$\epsilon_\textrm{SM}$ of the form
\begin{equation}
F_\textrm{SM} = \half K_2 \epsilon_\textrm{SM}^\beta
\end{equation}
where $K_2$ is the corresponding shear modulus for this mode. In the
case of the semi-soft Sm-$C$ elastomer, this term arises because of
the rapid rotation of the director during the start of the soft mode.

\begin{figure}[!htb]
\begin{center}
\includegraphics[width = 0.48\textwidth]{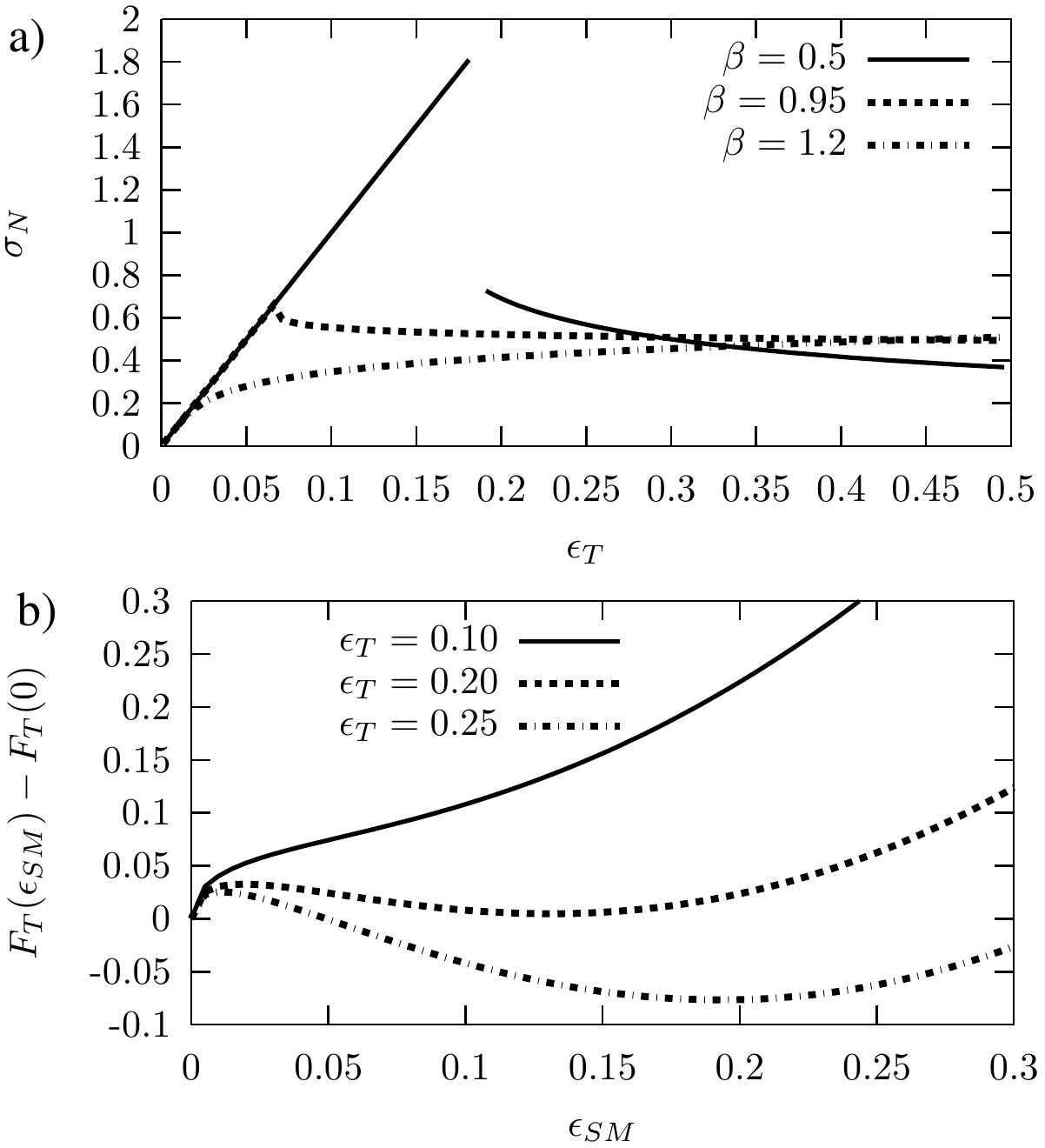}
\end{center}
\caption{a) An illustration of a discontinuous stress-strain curve for
  the scalar model described in the text, b) the free energy as a
  function of the variable $\epsilon_\textrm{SM}$ for fixed total
  strain values. Here $K_1 =10 $ and $K_2 = 1 $.  }
\label{fig:toydiscont}
\end{figure}

The total free energy is then
\begin{equation}
  F_\textrm{T} = \half K (\epsilon_\textrm{T} - \epsilon_\textrm{SM})^2 
+ \half K_2  \epsilon_\textrm{SM}^\beta,
\end{equation}
where first spring in this system is hookean, and the second is
non-linear, being infinitely stiff at zero strain for $0<\beta<1$, but
softening rapidly as strain increases. This should be minimised over
$\epsilon_\textrm{SM}$ to determine the distribution of strain between
the two springs. It can be solved analytically for $\beta = 0.5$. The
behaviour of this model is illustrated in
Fig.~\ref{fig:toydiscont}. For small $\beta$ this system has a
discontinuity in the stress-strain curve, but as $\beta$ is increased
the stress-strain response becomes continuous.
The free energy as a function of $\epsilon_\textrm{SM}$ is also
illustrated in Fig.~\ref{fig:toydiscont}. For small values of
$\epsilon_T$ there is only one minimum at $\epsilon_\textrm{SM} = 0$,
corresponding to no strain of the second spring. However, as the total
strain increases, the second mode of deformation becomes activated and
there is a minimum for larger values of $\epsilon_\textrm{SM}$. Since
there is a barrier between the two minima, the transition is first
order, so there is a jump in the equilibrium value of
$\epsilon_\textrm{SM}$. For larger values of $\beta$ the phase
transition becomes continuous, and the stress-strain curve no longer
exhibits a jump.

This behaviour is analogous to that of the semi-soft Sm-$C$ elastomer
as the free energy exhibits a discontinuity when stretched
perpendicular to the director (where the soft mode has a singular
edge). Larger values of $\beta$ correspond to stretching at a larger
angle to the director, where the soft mode does not have such a rapid
rotation of the director, and a corresponding sharp drop in the
lateral dimension. If the angle between the director and the
elongation direction is large enough, then the stress-strain response
becomes continuous as we will see in the next section.

\subsection{Elongation at an angle $\psi$ to the layer normal}

\begin{figure}[!htb]
\begin{center}
\includegraphics[width = 0.48\textwidth]{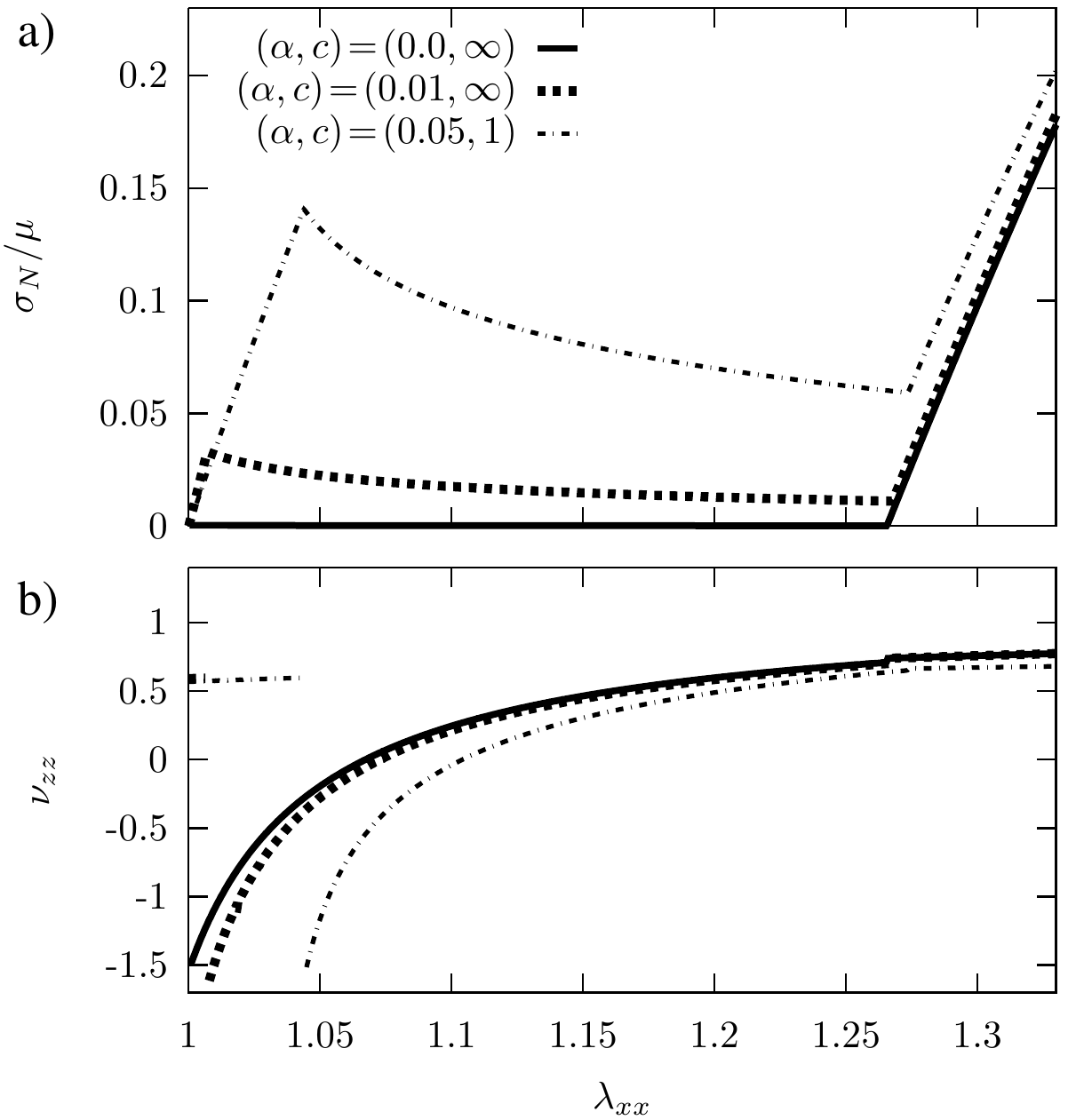}
\end{center}
\caption{a) The stress-strain curves for stretching at an angle of
  $\psi = 0.65$ radians to the layer normal, for $b = 60, r=2,
  \theta_0 = 0.5$ radians and various parameter values $(\alpha, c)$,
  and b) the Poisson's ratio in this geometry. }
\label{fig:sskqc}
\end{figure}

\begin{figure}[!htb]
\begin{center}
\includegraphics[width = 0.48\textwidth]{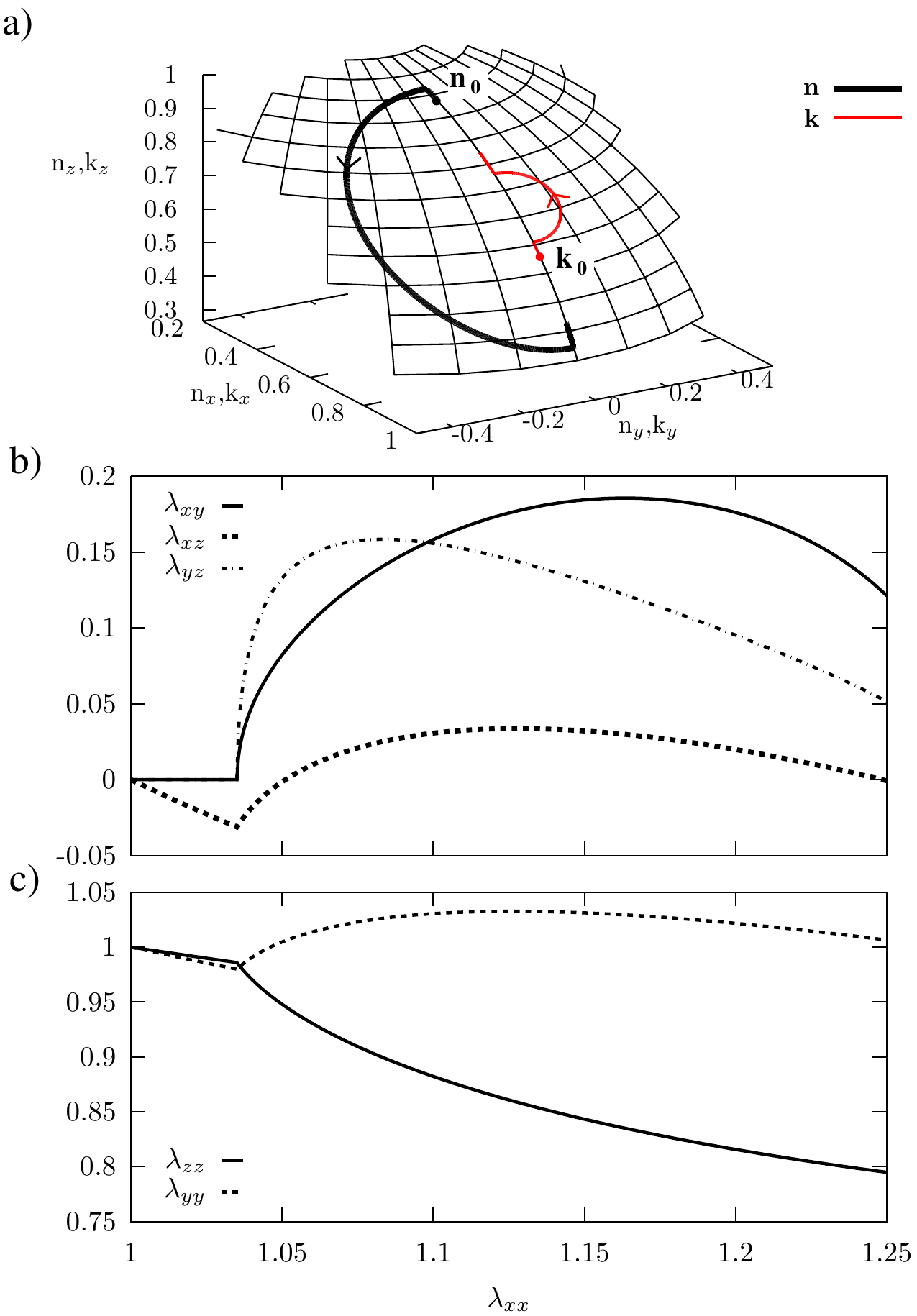}
\end{center}
\caption{For stretching at an angle of $\psi = 0.65$ radians to the
  layer normal a) shows the director and layer normal components in
  the $xy$ plane, b) the shear components of the deformation, and c)
  the diagonal components of the deformation for stretching at an
  angle of $\psi = 0.65$ radians to the layer normal, for the case
  $(\alpha, c) = (0.05, \infty)$, and $b = 60, r=2, \theta_0=0.5$. }
\label{fig:kqcdef}
\end{figure}

The last deformation we consider is shown in fig.~\ref{fig:geom}
d). The numerical solution of stress-strain curve associated with this
geometry is shown in Fig.~\ref{fig:sskqc}.
The stress-strain curve is continuous in this geometry, but again has
a pronounced negative slope. There is a negative IPR of $\sim -1.5$
that is roughly independent of the semi-soft parameter. The expansion
of the sample that accompanies the rotation of the director can be
seen in Fig.~\ref{fig:kqcdef}.

\section{Discussion}
\label{sec:discussion}

The first three deformations considered above in Fig.~\ref{fig:geom}
(a-c), when made with clamped boundary conditions, would not be soft
even without the semi-soft elastic term. This is because no
microstructure can be constructed from the soft deformations that is
compatible with the boundary conditions, due to the shear components
in the Sm-$C$ soft mode \cite{adams:07}. However, the properties of a
long sheet of Sm-$C$ LCE may approximate this behaviour as the centre
of the sample could deform without rigid boundary conditions. The
final deformation in Fig.~\ref{fig:geom} d) can be performed with
clamped boundary conditions in the soft case. In the semi-soft case
the sample starts to shear before the onset of rotation, which is not
compatible with clamped boundaries, so in experiment it may be even
stiffer initially due to this additional constraint on its
deformation.

The maximum lateral expansion can be deduced from the soft mode
presented in \S \ref{sec:softmodes}. The shear components are
transformed, through a rotation, into an elongation. At $\phi = \pi/2$
the maximum lateral expansion occurs (in the $y$ direction for the
example given in the text), and has a value of $\sqrt{\frac{r}{\rho}}$.

The region of negative slope in the constitutive models reported here
is typically explained by a Maxwell Construction. Similar behaviour
occurs in the Van der Waals gas model which has a region of negative
slope in the pressure-volume curve. Here there is a two phase region
consisting of a mixture of the liquid and gas phases. In solids the
two deformations on either side of the instability must be compatible
to form a mixture \cite{PhysRevE.78.011703}. The system should then
disproportionate, adopting a mixture of the two deformations to
achieve the externally imposed strain. The first order type phase
transition seen in the example stretching perpendicular to the layer
normal can result in hysteretic behaviour as the system jumps from one
energy well to another. The rate of the deformation in comparison to
the sample relaxation times may also result in hysteresis
\cite{1981agarwal}. There is interest in negative stiffness materials
\cite{lakesnature} for applications such as sealants, stiffening
composites, and creating meta-materials having a negative refractive
index to sound waves.

Experimental work reporting mechanical testing on Sm-$C$ monodomains
has not been reported. Whilst it is anticipated that these monodomains
should exhibit soft elasticity, the addition of the semi-soft
elasticity term to the model suggests that these effects may be
difficult to observe for large semi-soft parameter $\alpha$. When
stretching perpendicular to both the layer normal and the director,
the semi-soft term may prevent any stress plateau being observed,
instead only a shoulder is visible in the stress-strain response.

\begin{figure}[!htb]
\includegraphics[width = 0.48\textwidth]{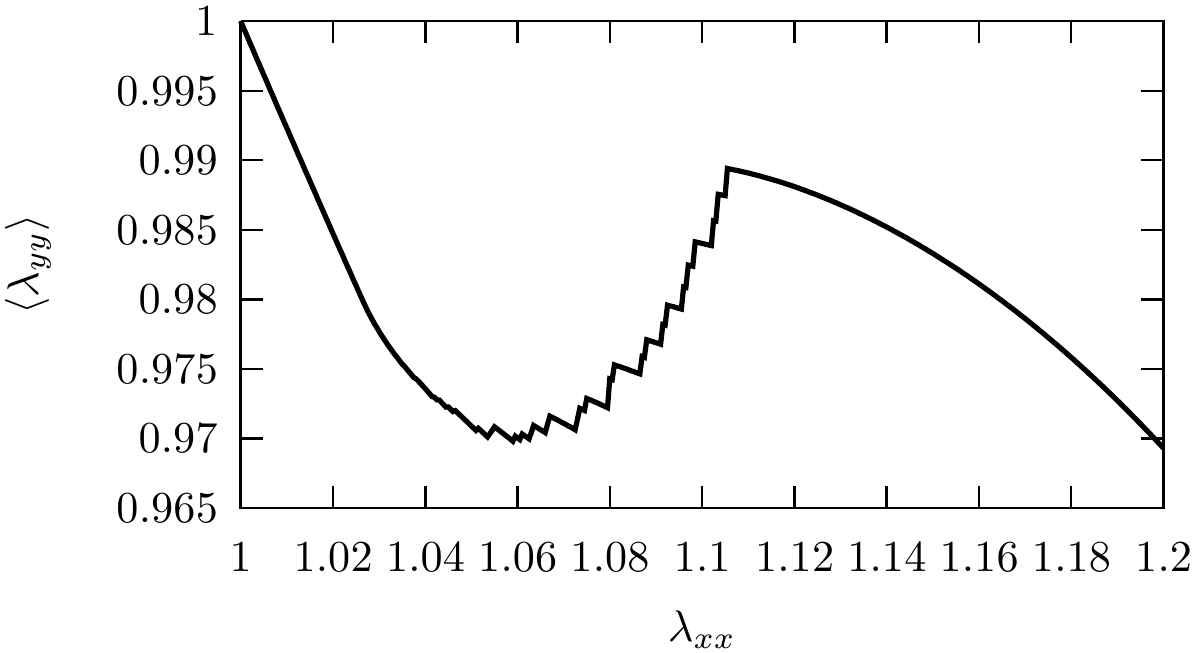}
\caption{The average value of $\lambda_{yy}$ for $50$ domains in a
  pseudo-monodomain illustrated in fig.~\ref{fig:pseudomd} as a
  function of $\lambda_{xx}$ assuming they all experience the same
  strain, and deform independently. Model parameters are $b=60, r=2,
  \theta_0 = 0.5, c = \infty, \alpha = 0.05$. }
\label{fig:polyyy}
\end{figure}

Although we have only considered the deformations of monodomains here,
the results inform model predictions for polydomains. Polydomains are
difficult to model because of the requirement of ensuring that
adjacent domains deform in a compatible way. A simplifying
approximation used to model a polydomain is to assume that it consists
of an array of monodomains that deform at the imposed external strain,
but are independent from each other. If we deform the pseudo-monodomain
shown in Fig.~\ref{fig:pseudomd} by stretching in the $\vec{x}$
direction, then deformation component $\lambda_{yy}$ averaged over all
the domains is illustrated in fig.~\ref{fig:polyyy} for $50$ domains.
This figure shows that there is a negative IPR as the director in each
of the domains jumps causing them to expand. The curve illustrated
here is jagged because the alignment of each domain jumps at a
slightly different threshold. The expansion of the film thickness, and
the energy loss as a result of the jump in the director orientation in
this geometry may be observable in experiments on pseudo-monodomains
\cite{ferrer2008,2008ren}. The larger values of deformation reported
in experiment before the knee in the stress-strain curve point to a
much larger value of $\alpha$ than in the illustrative plot in
Fig.~\ref{fig:polyyy}.

The features of the smectic-$C$ model described here would be present
in a wide range of models that have soft modes of nematic elastomers
but incorporate the constraint on the director to remain at a fixed
angle to the layer normal. However, validation of these models await
either experimental work on mechanical testing of Sm-$C$ monodomains,
or theoretical work on pseudo-monodomains to link up with existing
mechanical experiments on pseudo-mododomains. 

\section{Conclusion}

We have studied a model of monodomain Sm-$C$ LCEs with the inclusion
of a semi-soft elastic term to describe imperfections in the
elastomer.  As result of the negative incremental Poisson's ratio
inherent in the soft modes of a Sm-$C$ monodomain, the mechanical
properties of a semi-soft monodomain are unusual. When stretching
perpendicular to the layer normal and the director, the response is
reminiscent of a nematic elastomer. A finite force is required to
deform the LCE and initial the rotation of the director. However, the
stress plateau is less well defined for larger values of semi-soft
parameter $\alpha$; it is reduced to a shoulder in the stress-strain
response. When stretching parallel to the layer normal the elastomer
again exhibits a threshold to director rotation. Once director
rotation has started the elastomer has a negative incremental
Poisson's ratio, and a negative stiffness. A negative incremental
Poisson's ratio of up to $\nu\sim-1.5$ has been found for typical
model parameters. This arises because the director rotates in a
direction perpendicular to the stretch axis due to the constraint of
the layer normal. This more detailed understanding of monodomain
deformations of Sm-$C$ elastomers might prove useful in understanding
recent mechanical and piezoelectric experiments on polydomain Sm-$C$
elastomers.

\begin{acknowledgements}
  We would like to thank SEPnet for supporting for this project, and
  Dr Daniel Corbett and Dr James Busfield for helpful discussions.
\end{acknowledgements}

\appendix

\section{Transforming the soft mode for different starting
  configurations}
\label{app:smlayernormal}

The soft mode given by Eq.~(\ref{eqn:softmode}) can be transformed to
other geometries by a pair of rotation matrices. For example, consider
the case of a Sm-$C$ elastomer stretched parallel to the layer
normal. Let us assume that starting layer normal is $\vec{k}_0 =
\vec{x}$ and the starting director is $\vec{n}_0 = \cos\theta_0
\vec{x} + \sin \theta_0 \vec{y}$. The soft mode for this configuration
that is an upper triangular matrix, as described in
Eq.~(\ref{eqn:defform}), can be found as follows. From the reference
configuration a body rotation is performed such that the layer normal
$\vec{k}_0$ is parallel to the $\vec{z}$ axis. In this case, a
$90^\circ$ rotation about the $\vec{y}$ axis
\begin{equation}
\ten{Q} = \left( 
\begin{array}{ccc}
0&0&-1\\0&1&0\\1&0&0
\end{array}
\right).
\end{equation}
After this rotation the director is given by $\vec{n} = \vec{z} \cos
\theta_0 + \vec{y} \sin \theta_0$. Note that in general an additional
rotation around the $\vec{z}$ axis may be required to ensure the
director is in this orientation. This is the initial configuration for
the soft mode given in Eq.~(\ref{eqn:softmode}). The director now
rotates by an angle $\phi$ around the new layer normal, and the sample
executes the soft mode. Finally a rotation of the target state is
performed such that the deformation matrix has the form described in
Eq.~(\ref{eqn:defform}). This rotation matrix is in general simpler if
we first undo the rotation $\ten{Q}$. The rotation matrix $\ten{P}$ is
described by three angles:
\begin{eqnarray}
\nonumber
\ten{P}&=& \left( 
\begin{array}{ccc}
\cos \psi_z & \sin \psi_z & 0\\ -\sin \psi_z & \cos \psi_z & 0 \\ 0&0&1
\end{array}
\right)\cdot
\left( 
\begin{array}{ccc}
1&0&0\\0&\cos \psi_x & \sin \psi_x \\0& -\sin \psi_x & \cos \psi_x 
\end{array}
\right)\\&
\cdot&
\left( 
\begin{array}{ccc}
\cos \psi_y & 0&\sin \psi_y \\0&1&0\\ -\sin \psi_y &0& \cos \psi_y 
\end{array}
\right).
\end{eqnarray}
The three angles $\psi_x, \psi_y$ and $\psi_z$ can be calculated by
substituting into the equation
\begin{equation}
\ten{\lambda} = \ten{P} \cdot \ten{Q}^T\cdot \ten{\lambda}_\textrm{soft} \cdot \ten{Q}
\end{equation}
and ensuring that the three lower triangular elements of
$\ten{\lambda}$ are zero. In the case of stretching parallel to
$\vec{k}_0$ and perpendicular to $\vec{n}_0$ the soft mode can be
calculated analytically. The algebraic expressions for these soft
modes is long, and unedifying so will not be presented here.

\bibliographystyle{apsrevM}  
\ifx\mcitethebibliography\mciteundefinedmacro
\PackageError{apsrevM.bst}{mciteplus.sty has not been loaded}
{This bibstyle requires the use of the mciteplus package.}\fi

\end{document}